\documentclass{article}
\usepackage{arxiv}

\usepackage{cite}
\usepackage{amsmath,amsfonts,bm}
\usepackage{array}
\usepackage[caption=false,font=small]{subfig} 
\usepackage{subfig}
\usepackage{textcomp,stfloats,url,verbatim,graphicx}

\usepackage{subfiles} % subfile
\usepackage{booktabs,tabularx,multirow}

\usepackage[ruled]{algorithm2e}
\usepackage{setspace}
\SetKwInput{KwInput}{Input}
\SetKwComment{Comment}{$\triangleright \;$}{}

\graphicspath{{figures/}}
\labelformat{subfigure}{\thefigure(\emph{#1})}

\begin{document}

\title{Traffic Signal Cycle Control with Centralized Critic and Decentralized Actors under Varying Intervention Frequencies}

\author{Maonan Wang \textsuperscript{\textsection} \\
	The Chinese University of Hong Kong, Shenzhen, China \\
    Shanghai AI Laboratory, Shanghai, China \\
	\texttt{maonanwang@link.cuhk.edu.cn} \\
	\And
	Yirong Chen \textsuperscript{\textsection} \\
	Stanford University, California, USA \\
	\texttt{chenyr@stanford.edu} \\
	\AND
	Yuheng Kan \\
	SenseTime Group Limited, Shanghai, China \\
	Shanghai AI Laboratory, Shanghai, China \\
	\texttt{kanyuheng@sensetime.com} \\
	\And
	Chengcheng Xu \\
	SenseTime Group Limited, Shanghai, China \\
	\texttt{xuchengcheng@sensetime.com} \\
	\And
	Michael Lepech \\
	Stanford University, California, USA \\
	\texttt{mlepech@stanford.edu} \\
	\And
	Man-On Pun \\
	The Chinese University of Hong Kong, Shenzhen, China \\
	\texttt{simonpun@cuhk.edu.cn} \\
	\And
	Xi Xiong \\
	Tongji University, Shanghai, China \\
	\texttt{xi\_xiong@tongji.edu.cn} \\
}

\maketitle
\renewcommand{\shorttitle}{Traffic Signal Cycle Control under Varying Intervention Frequencies}
\begingroup\renewcommand\thefootnote{\textsection}
    \footnotetext{Equal contribution}
\endgroup

% %%%%%%%%
% Abstract
% %%%%%%%%
\begin{abstract}
Traffic congestion in urban areas is a significant problem, leading to prolonged travel times, reduced efficiency, and increased environmental concerns. Effective traffic signal control (TSC) is a key strategy for reducing congestion. Unlike most TSC systems that rely on high-frequency control, this study introduces an innovative joint phase traffic signal cycle control method that operates effectively with varying control intervals. Our method features an \textit{adjust all phases} action design, enabling simultaneous phase changes within the signal cycle, which fosters both immediate stability and sustained TSC effectiveness, especially at lower frequencies. The approach also integrates decentralized actors to handle the complexity of the action space, with a centralized critic to ensure coordinated phase adjusting. Extensive testing on both synthetic and real-world data across different intersection types and signal setups shows that our method significantly outperforms other popular techniques, particularly at high control intervals. Case studies of policies derived from traffic data further illustrate the robustness and reliability of our proposed method.
\end{abstract}


\maketitle

% %%%%%%%%%%%%
% Introduction
% %%%%%%%%%%%%
\section{Introduction} \label{sec_introduction}

% Why TSC is important & conventional TSC methods
Road traffic congestion is a global issue that results in environmental degradation, economic losses, and diminished quality of life for commuters. According to the Global Traffic Scorecard developed by INRIX, traffic congestion costs an average American around $99$ hours per year and around $\$1,400$ per year in lost economic activities based on 2019 data \cite{INRIX}, not to mention the environmental impacts caused by excessive emissions. One of the most effective ways to reduce traffic congestion is through optimal traffic signal control (TSC) \cite{aslani2017adaptive,liu2017distributed,haydari2020deep,du2021coupled,su2022emvlight}. By regulating traffic light timing, TSC can ease congestion and improve traffic flow. Various approaches for optimal traffic signal control have been proposed, such as the Webster model \cite{webster1958traffic} and the HCM delay model \cite{manual2000highway}. However, these methods often rely heavily on simplified traffic models or make unrealistic assumptions of a uniform arrival rate of vehicles. Existing TSC systems such as SCOOT \cite{hunt1982scoot} and SCATS \cite{lowrie1990scats} also face challenges in managing complex traffic scenarios.

% RL-based methods 
To address these challenges, deep reinforcement learning (RL) has emerged as a promising tool for optimal TSC systems. RL allows the traffic signal controller to learn and optimize signal timings in real-time based on the current traffic conditions. RL-based TSC systems have demonstrated significantly improved traffic flow and reduced congestion as compared to conventional methods in several studies \cite{wei2019survey, chu2019multi, haydari2020deep, li2020multi, noaeen2022reinforcement, wang2023unitsa, gu2024pi}. By directly interacting with the environment, the RL agent learns a superior signal plan by analyzing thousands of samples that reflect traffic state changes and action choices.

% Four action designs in RL-based TSC
To reduce congestion, existing RL-based algorithms primarily focus on modifying the duration of phases individually. These studies typically employ one of the four following action designs: 
1) \textit{choose next phase} \cite{wei2019presslight,gu2024pi}, Selecting a phase from all possible phases at each time step. % choose next phase
2) \textit{next or not} \cite{li2016traffic,wang2023unitsa}, determining whether to switch to the next phase or stay on the current one at each time step. % next or not
3) \textit{set current phase duration} \cite{aslani2017adaptive}, setting the phase duration at the beginning of each phase. % set current phase duration
4) \textit{adjust single phase} \cite{liang2019deep}, modifying only one phase in the whole cycle. % adjust single phase
Particularly, the action of \textit{choose next phase} is most widely used by recent studies for both single intersection settings and large-scale city networks due to its flexibility \cite{wei2019survey}. By allowing the agent to select any phase at each time step, this approach acts more frequently than other action designs, and thus is more likely to converge to an optimal policy. Section~\ref{related_RL} will delve deeper into the four action designs, discussing their effectiveness and shortcomings in detail. 

% Two limitations
Despite significant advancements in traffic efficiency through RL-based TSC methods, existing research has not adequately addressed certain practical constraints for real-world application. One such oversight is the intervention frequency, the interval between consecutive adjustments made by the TSC system to the traffic signals. This aspect is critical in real-world applications due to \textit{limited resources}, \textit{safety concerns}, \textit{disruption to traffic flow}, and \textit{stability of the system}. 1) \textit{Limited resources}, in some cases, TSC systems may have limited resources such as processing power or network bandwidth \cite{xing2022tinylight}. Adjusting the traffic signal too frequently may not be feasible or may result in degraded performance due to resource constraints. 2) \textit{Safety concerns, overly frequent signal changes could confuse drivers, potentially leading to accidents. Moreover, a lower frequency of intervention may align better with the need for manual verification, a common practice in fields that require cautious AI deployment, such as medicine and finance} \cite{bhardwaj2017study, rajpurkar2022ai, du2019fridays}. 3) \textit{Disruption to traffic flow}, rapid changes applied to traffic lights can disrupt traffic flow. Stable changes are vital for route planning and vehicle platooning in urban traffic management \cite{xiong2021optimizing, cao2022book}. 4) \textit{Stability of the TSC system}, frequent modifications to the traffic signal can result in oscillations or other instabilities, negatively affecting the overall performance of the system \cite{li2017evaluating}. The appropriate intervention frequency depends on the specific circumstances and constraints of the traffic control system.

To address the limitations of existing approaches, this study introduces a novel joint phase traffic signal cycle control strategy. This approach employs a \textbf{C}entralized \textbf{C}ritic and \textbf{D}ecentralized \textbf{A}ctors (CCDA) framework, taking into account the critical factor of intervention frequency. Our approach also introduces a new action design termed \textit{adjust all phases}, which alters the duration of all phases during a signal cycle. By adjusting all phases with each decision-making action, we enhance the effectiveness of each intervention. While the \textit{adjust all phases} concept is beneficial, it can dramatically increase the action space with a centralized training approach. To mitigate this, our method employs the CCDA architecture, where decentralized actors adjust individual signal phases, substantially narrowing the action space. Concurrently, a centralized critic assesses the complete traffic situation, facilitating the synchronization among the various actors. This ensures a smooth traffic flow and maximizes the overall benefit to the system. By modifying all traffic phases for each action and minimizing the action space through the CCDA framework, our approach does not target a single optimal frequency. Instead, it focuses on the adaptability of the control system across various frequencies.

The contributions of this work can be summarized as follows: 
\begin{itemize}
    \item The design of the proposed RL-based TSC explicitly incorporates intervention frequency. To maximize the effectiveness of each interaction with the traffic environment, we introduce a novel action design, \textit{adjust all phases}, allowing for simultaneous fine-tuning across all traffic signal phases.
    \item We present an adaptive control method termed CCDA. In this framework, decentralized actors handle the complex action space, while the centralized critic facilitates coordination among the traffic signal phases, promoting overall system efficiency.
    \item Extensive testing using both synthetic and real-world datasets has shown that our approach outperforms existing methods across various intervention frequencies. For instance, at low control frequencies, our method reduces the average queue length by up to $58.1\%$ compared to other RL-based methods, while also enhancing signal stability. Additionally, we provide a detailed analysis of the policies derived from these methods~\footnote{Our simulation code is open-sourced and available at \url{https://github.com/Traffic-Alpha/CCDA-Light}}.
\end{itemize}

% Paper organization
The remainder of the paper is organized as follows: Section~\ref{sec_related_work} presents the literature review, while Section~\ref{sec_preliminary} defines the concept of intervention frequency. Section~\ref{sec_method} then details the proposed CCDA approach, including the agent design and novel actor-critic approach. Following that, Section~\ref{sec_experiment} presents the experiment setup and compared methods. Finally, Section~\ref{sec_results} discusses and analyzes the performance of the proposed framework as compared to other algorithms under various intervention frequencies before Section~\ref{sec_conclusion} concludes the paper.

% %%%%%%%%%%%%
% Related Work
% %%%%%%%%%%%%
\section{Related work} \label{sec_related_work}

% Conventional TSC
\subsection{Conventional Traffic Signal Control}

The optimization of traffic signals in urban areas has long been a persistent challenge in reducing congestion. Several classical methods based on rules or mathematical models have been developed to optimize the traffic signal in response to diverse scenarios. For instance, the Webster method \cite{webster1958traffic} calculates the optimal cycle length and phase split for a given intersection based on traffic volume, assuming a uniform traffic flow over a certain period (e.g., $10$ minutes). Furthermore, \cite{gershenson2004self} established the Self-Organizing Traffic Light Control (SOTL), which uses rules to determine whether to keep or change the current phase. Specifically, when the number of vehicles approaching the green signal side exceeds a preset threshold, the SOTL method requests the traffic light to remain in its current phase; Otherwise, the traffic light changes to the next phase. 

There are also two successful commercial TSC systems, namely the Split Cycle Offset Optimization Technique (SCOOT) \cite{hunt1982scoot} and the Sydney Coordinated Adaptive Traffic System (SCATS) \cite{lowrie1990scats}. These TSC systems have been installed at more than $50,000$ intersections in over $180$ cities. These systems choose cycle length, phase split, and offset from several pre-defined plans according to the data derived from road traffic sensors (i.e., loop detectors and vehicle tracking cameras). While these conventional methods have demonstrated some efficacy in alleviating traffic congestion, they fall short of fully leveraging real-time traffic information. Furthermore, these methods often rely on simplified assumptions or require expertise to determine essential traffic parameters, such as saturation flow and distance headway \cite{mathew2007fundamental}.

\begin{figure*}[!ht]
    % Four common action designs in TSC
    \centering
    \subfloat[]{\includegraphics[width=0.45\textwidth]{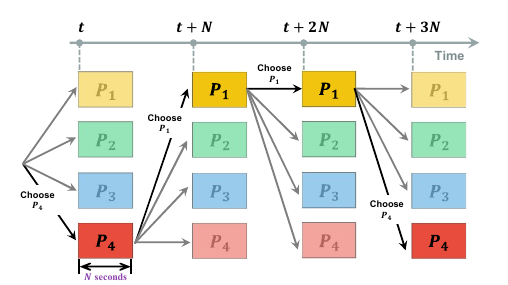}%
    \label{fig_choose_next_phase}}
    \hfil
    \subfloat[]{\includegraphics[width=0.45\textwidth]{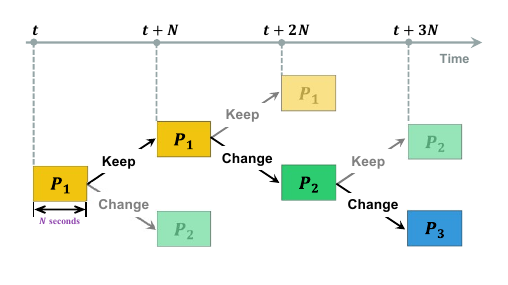}%
    \label{fig_next_or_not}}
    
    \hfill
    
    \subfloat[]{\includegraphics[width=0.45\textwidth]{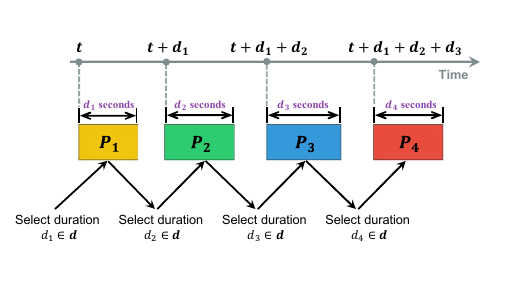}%
    \label{fig_set_phase_duration}}
    \hfil
    \subfloat[]{\includegraphics[width=0.45\textwidth]{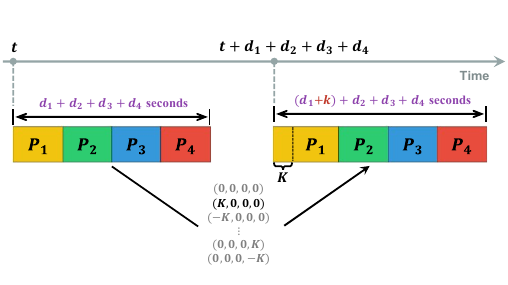}%
    \label{fig_adjust_single_phase}}
    \caption{Four commonly used action designs in the RL-based TSC problem. (a) Choose next phase. (b) Next or not. (c) Set current phase duration. (d) Adjust single phase.}
    \label{fig_four_action_designs}
\end{figure*}

% RL-based TSC
\subsection{RL-based Traffic Signal Control} \label{related_RL}

In recent years, RL-based TSC systems have demonstrated efficacy in alleviating congestion. 
Among the proposed systems, there are four commonly used action designs, namely \textit{choose next phase}, \textit{next or not}, \textit{set current phase duration} and \textit{adjust single phase}. Fig.~\ref{fig_four_action_designs} illustrates examples of these four action designs for a TSC system with four phases, assuming that $t$ is the starting time.

\textit{Choose next phase} is the most commonly used action design in the RL-based TSC problem \cite{wei2019presslight, wei2019colight, lee2019reinforcement, oroojlooy2020attendlight, chen2020toward, wang2021gan, boukerche2021novel, shabestary2022adaptive, ghanadbashi2022using}. In this design, the system selects the next phase among all possible phases every $N$ seconds. As illustrated in Fig.~\ref{fig_choose_next_phase}, the agent selects $P_{4}$ at time $t$, which lasts for $N$ seconds. After $N$ seconds, the agent chooses $P_{1}$ based on the observation at the intersection. At the time $t+2N$, the agent selects $P_{1}$ once more, which similarly lasts for $N$ seconds. At the time $t+3N$, the agent takes action $P_{4}$. As shown in Fig.~\ref{fig_choose_next_phase}, it can be observed that the phase sequence is disrupted with this action design. Despite its flexibility, this design results in a random phase sequence which can pose safety risks to traffic participants. Drivers and pedestrians typically have expectations about the signal sequence, such as anticipating a straight-through green light for north-south traffic will be followed by a left-turn phase. As a result, such unpredictability in signal changes can be confusing and uncomfortable for traffic participants if the signal plan does not meet their expectations.

To maintain a cyclical change in traffic signals, some studies employ the \textit{next or not} action design \cite{li2016traffic, mannion2016experimental, van2016coordinated, wei2018intellilight,lin2018efficient,paul2022exploring}. In this design, the agent decides every $N$ seconds whether to maintain the current phase or switch to the next phase. As depicted in Fig.~\ref{fig_next_or_not}, the traffic signal is in $P_{1}$ at time $t$. After $N$ seconds, the agent takes action ``keep'', maintaining phase $P_{1}$. At time $t+2N$, the agent chooses the action ``change'', switching the phase from from $P_{1}$ to $P_{2}$. It is evident that this action design is capable of maintaining the phase sequence. 

The third action design used in RL-based TSC problem is known as \textit{set current phase duration} \cite{aslani2017adaptive, aslani2018traffic, wang2022adlight}. In this design, the agent learns to determine the length of the current phase by selecting from a predetermined set of time intervals. As shown in Fig.~\ref{fig_set_phase_duration}, the agent sets the duration of $P_{1}$ to be $d_{1}$ based on the information gathered from the intersection at time $t$. After $d_{1}$ seconds, the agent sets the duration of phase $P_{2}$ to be $d_{2}$. This action design modifies only the duration of each phase, thereby preserving the cyclical sequence of the phases. 

The three action designs mentioned above do not determine traffic signal plans in cycles, even though signal schemes are typically updated in cycles \cite{chiu1992adaptive}. To address this, Liang \textit{et~al.} \cite{liang2019deep} proposed a new action design named \textit{adjust single phase}, where the agent selects one phase from the traffic signal plan and adjusts its duration once per cycle. To ensure smooth transitions between actions, the change in each phase duration is limited to a maximum of five seconds in \cite{liang2019deep}. As illustrated in Fig.~\ref{fig_adjust_single_phase}, if the phase durations at time $t$ are $(d_{1}, d_{2}, d_{3}, d_{4})$, the agent's action $(k,0,0,0)$ after a full cycle adds $k$ seconds to $P_{1}$, resulting in new durations of $(d_{1}+k, d_{2}, d_{3}, d_{4})$. While this design promotes gradual signal changes by adjusting a single phase, it may not perform well in complex traffic situations or with infrequent interventions, such as multi-phase intersections or in rapidly changing traffic conditions. In contrast, our proposed action design, \textit{adjust all phases}, is more adept at handling such scenarios by allowing simultaneous adjustments to all phases.

Furthermore, our research extends previous studies by examining how different intervention frequencies impact TSC system performance, a significant factor that has not received enough attention. Only a limited number of studies have considered the \textit{frequency of signal change} by developing reward functions that discourage frequent signal changes \cite{van2016coordinated, wei2018intellilight}. These functions penalize the agent for each signal alteration, but accurately calibrating the penalty is challenging since small adjustments can lead to large-scale effects on system behavior. Additionally, due to the limitations of their agent designs, these methods do not perform well under conditions of low-frequency interventions. Our study addresses this by incorporating intervention frequency as a key factor in the agent's interaction with the traffic environment. By embedding this consideration into our CCDA method and introducing a new action design, our approach balances stability and efficiency, accommodating a diverse range of intervention frequencies across various operational conditions.

Quantifying the performance of different TSC systems is also a vital task. Various measures have been proposed, such as (1) the average travel time of vehicles in the network, (2) the queue length in the road network, (3) the number of stops that vehicles make in the network, and (4) the number of vehicles that complete their trip during a period (throughput) \cite{wei2019survey}. However, these evaluation measures neglect to examine the traffic control policies learned by the RL agent. In this study, we emphasize the significance of investigating the control policies rather than solely focusing on traffic efficiency. Our experiments showcase several notable policies learned from both simulated and real traffic flow.

% %%%%%%%%%%%%
% Preliminary
% %%%%%%%%%%%%
\begin{figure}[!htbp]
  \centering
  % movement
  \subfloat[]{\includegraphics[width=0.4\linewidth]{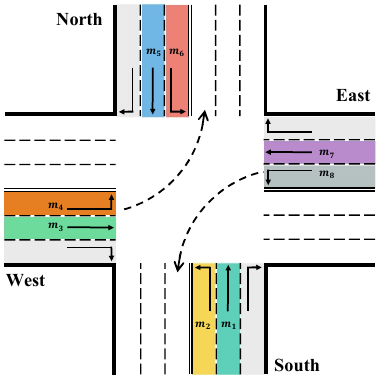}
  \label{fig_movement}}
  
  % phase
  \subfloat[]{\includegraphics[width=0.4\linewidth]{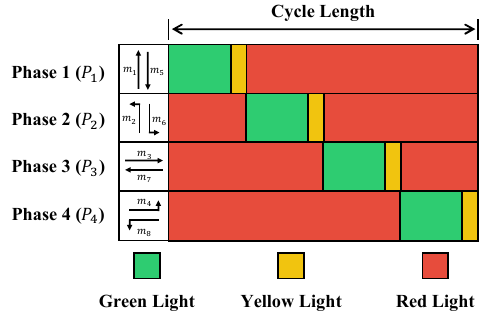}
  \label{fig_phase}}
  
  \caption{A standard 4-way intersection with its movements, phases, and cycle. (a) Topology of the intersection; (b) Relationship between the four traffic phases and the cycle.}
  \label{fig_term_defination}
\end{figure}

\section{Preliminaries} \label{sec_preliminary}

In this section, we define key terminology used in the field of TSC to aid in problem formulation.

% Movement, phase and cycle
\subsection{Traffic Movement, Signal Phase, and Cycle}
A traffic movement refers to a connection between an incoming lane to an outgoing lane. For illustration, consider a standard 4-way intersection as shown in Fig.~\ref{fig_movement}. At this intersection, there are eight movements, denoted as $\{m_i\}$ for $i=1,2,\ldots, 8$. It is important to note that right turns are not considered in this study, as it is assumed they are always permitted with caution.

A traffic phase is defined as the control over one or more traffic movements. A traffic cycle represents the total time required for a complete sequence of all signal phases at an intersection. By managing the durations or sequences of these phases, traffic engineers can effectively control traffic flow and mitigate congestion. Fig.~\ref{fig_phase} illustrates the four phases made up of the eight movements mentioned. These phases are: north-south ($P_{1}=\{m_{1},m_{5}\}$), north-east and south-west ($P_{2}=\{m_{2},m_{6}\}$), east-west ($P_{3}=\{m_{3},m_{7}\}$), and east-south and west-north ($P_{4}=\{m_{4},m_{8}\}$). The green grid in Fig.~\ref{fig_phase} shows the duration of the green light for each phase, indicating that $P_{i+1}$ turns green following the red signal of $P_{i}$. The cycle length is the period from the start of $P_{1}$ to the end of $P_{4}$.

% Intervention frequency
\subsection{Intervention frequency in TSC}

In RL-based TSC, intervention frequency refers to how often the agent adjusts the traffic signals in response to changes in traffic patterns. A high intervention frequency can increase the system's responsiveness to dynamic traffic conditions and enhance its ability to manage unforeseen events, such as accidents or sudden shifts in traffic flow. Conversely, a low intervention frequency might cause the control policy to become outdated, potentially leading to suboptimal performance. However, maintaining a lower frequency can prevent the system from overfitting to specific traffic conditions, thereby improving its ability to generalize across various traffic scenarios. This also tends to stabilize the control policy.

In practical settings, traffic engineers must select an appropriate intervention frequency, taking into account factors such as resource limitations and safety concerns. In this study, we use $\Delta t$ to represent the time interval that regulates intervention frequency. A larger $\Delta t$ indicates a lower frequency of interaction, while a smaller $\Delta t$ suggests a higher frequency.

\begin{figure}[!ht]
  \centering 
  \includegraphics[width=0.4\linewidth]{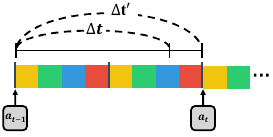}
  \caption{An example of applying the intervention frequency based on cycle-based control action design in a four-phase traffic signal system.}
  \label{fig_intervention_frequency}
\end{figure}

Fig.~\ref{fig_intervention_frequency} illustrates how intervention frequency is applied in a cycle-based control action design, such as \textit{adjust single phase}. After the agent executes action $a_{t-1}$, it waits for $\Delta t^{\prime}$ seconds before initiating another action $a_{t}$. During this interval, the traffic signal maintains the settings determined by action $a_{t-1}$. The value of $\Delta t^{\prime}$ is calculated using the following equation:
\begin{equation} \label{eq_intervention_frequency}
    \Delta t^{\prime} = \left\lceil \frac{\Delta t}{C} \right\rceil \times C,
\end{equation}
where $C$ denotes the cycle length of the traffic light following action $a_{t-1}$, and $\Delta t$ is the predefined intervention time interval. This calculation ensures that the agent's adjustments align with the beginning of each traffic light cycle. Notably, $\Delta t^{\prime}$ is always equal to or greater than $\Delta t$, depending on the specific cycle length $C$.

% %%%%%%%
% Method
% %%%%%%
\section{Methodology} \label{sec_method}

In this section, we detail the methodology of a centralized critic and decentralized actors with the \textit{adjust all phases} action design. We start by outlining the overall framework for joint phase traffic signal cycle control. Subsequently, we introduce a novel agent configuration that incorporates the \textit{adjust all phases} action design for phase adjustments. Following this, we describe the CCDA framework for traffic signal optimization, which integrates decentralized actors for individual traffic phases with a centralized evaluation mechanism for coordinated control. Finally, we present the proposed neural network architecture and the CCDA training workflow in detail.

\begin{figure}[!ht]
  \centering
  \includegraphics[width=0.6\linewidth]{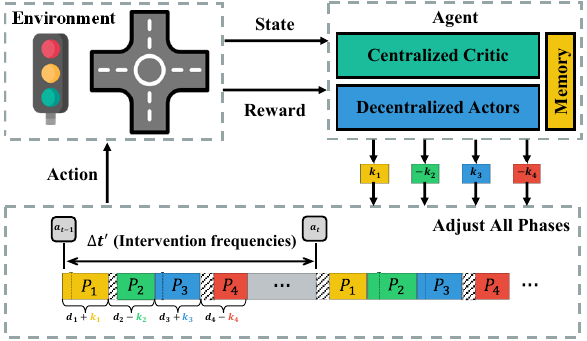}
  \caption{The framework of AAP (CCDA) with the intervention frequency.}
  \label{fig_introduction_framework}
\end{figure}

\subsection{Framework for Joint Phase Control} \label{framework}

Fig.~\ref{fig_introduction_framework} illustrates the overall structure of the proposed AAP (CCDA), which incorporates the concept of intervention frequency $\Delta t'$ in the TSC system design. We first extract three types of features that describe the traffic conditions: intersection vehicle information, road topology information, and traffic light information, which constitute the state $\mathbf{s}$. The improvement in traffic resulting from the actions taken is evaluated using the reward $r$, based on queue length (further details in Section~\ref{sec_agent_design}).

Subsequently, the agent takes action $\mathbf{a}$, which involves adjusting the durations of all phases simultaneously based on the state $\mathbf{s}$ from the environment. Given the use of decentralized actors, the action is represented as $\mathbf{a} = \{a^1, \ldots, a^N\}$, where $N$ is the total number of phases within a traffic signal cycle. The agent then waits for $\Delta t'$ seconds, which varies depending on the specific conditions and constraints of the TSC system. Following this, the agent observes the next state $\mathbf{s}'$ from the environment and stores the tuple $(\mathbf{s}, \mathbf{a}, r, \mathbf{s}')$ in the trajectory memory. The centralized critic evaluates the overall states $\mathbf{s}$ and $\mathbf{s}'$ to assess the outcomes of $\mathbf{a}$ and facilitate collaboration among the decentralized traffic phase actors (detailed in Section~\ref{ccda_framework}).

\subsection{Agent Configuration for Phase Adjustment} \label{sec_agent_design}

% State
\textbf{State}: For clarity, we focus on a single intersection in our discussion. For each traffic movement $i$ at the intersection, the state component $\mathbf{x}_i$ comprises three parts. The first part quantifies the congestion level of movement $i$, including the average traffic flow $F_i$, maximum occupancy $O^{max}_i$, and average occupancy $O^{avg}_i$ over the past $k$ seconds. These metrics can be readily obtained through traffic cameras or loop detectors. The second part includes two movement features: the direction of the movement $i$ (noting if it is a straight movement) $I^{s}_i$ and the number of lanes $L_i$ for the movement. The third part details the status of the traffic signal, capturing whether the movement is currently allowed (green signal) $I^{g}_i$, the duration of the current signal phase $D_i$, and whether this duration meets the minimum green light requirement $I^{min}_i$. The complete state $\mathbf{s}$ is defined as:

\begin{equation}
    \mathbf{s} = [\mathbf{x}_1, \mathbf{x}_2, \ldots, \mathbf{x}_8],
\end{equation}
where $\mathbf{x}_i = [F_i, O^{max}_i, O^{avg}_i, I^{s}_i, L_i, I^{g}_i, D_i, I^{min}_i]$. This configuration assumes that an intersection comprises 8 traffic movements. If an intersection has fewer than 8 movements, zero padding is used to maintain a consistent state representation \cite{wang2023unitsa}.

% Action
\textbf{Action}: An effective action design for TSC system must consider three key elements: (a) \textit{cycle-based management}: the design should manage traffic signals over cycles to accommodate driver behavior and pedestrian safety \cite{du2022safelight}; (b) \textit{steadiness}: the system must maintain stability by ensuring that the duration of each traffic signal phase does not vary significantly within a short period; (c) \textit{generalization ability}: the design should adapt to a broad range of traffic conditions, including various intersection configurations, traffic flows, and intervention frequencies.

\begin{figure}[!ht]
  \centering
  \includegraphics[width=0.6\linewidth]{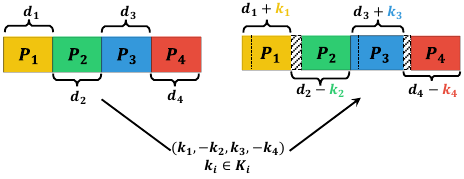}
  \caption{An example of \textit{adjust all phases} in a four phases traffic signal system.}
  \label{fig_adjust_all_phases}
\end{figure}

To address these requirements, this study proposes a novel action design, \textit{adjust all phases}, which simultaneously adjusts the durations of all phases at an intersection. Fig.~\ref{fig_adjust_all_phases} illustrates the application of \textit{adjust all phases} at a four-phase intersection. At the end of each cycle, the duration of each phase is adjusted; for example, the duration of phase 1 changes from $d_1$ seconds to $d_1 + k_1$ seconds, while the duration of phase 2 decreases from $d_2$ seconds to $d_2 - k_2$ seconds.

This action design maintains the original phase sequence and allows for adjustments to the traffic signals after completing each cycle. It is particularly effective in complex scenarios with frequent traffic flow changes or low intervention frequencies, as it enables the simultaneous adjustment of all phases. This contrasts with the \textit{adjust single phase} method, which modifies one phase at a time \cite{liang2019deep}. The proposed strategy ensures minimal variations in green light duration from one cycle to the next by applying only small adjustments (e.g., $\pm 3$ s or $\pm 5$ s) to each phase during each cycle, thereby maintaining the stability of the TSC system.

% Reward 
\textbf{Reward}: The average queue length of all lanes at the intersection is used as the reward, in place of conventional metrics such as waiting time or travel time delay. This choice is made because conventional metrics are hard to obtain from real-world traffic detection devices. With the queue length of the $j$-th lane in the $i$-th movement denoted by $q^{j}_{i}$, the reward $r$ is given by: 

\begin{equation} \label{eq_reward}
    r = - \sum_{i=1}^{8} \sum_{j=1}^{N_{i}}{q^{j}_{i}},
\end{equation}
where $N_{i}$ is the number of lanes for the $i$-th movement. The reward is normalized to prevent instability in the system, which could arise from excessively long queue lengths during traffic congestion.

\begin{figure}[!ht]
  \centering
  \includegraphics[width=0.5\linewidth]{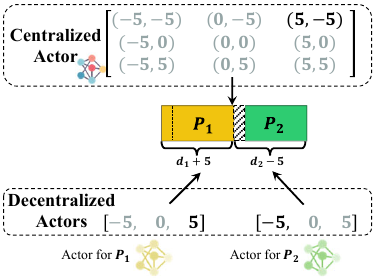}
  \caption{Comparison of centralized and decentralized actors in a two-phase traffic signal system.}
  \label{fig_centralized_decentralized}
\end{figure}

\subsection{CCDA Framework for Traffic Signal Optimization} \label{ccda_framework}

\subsubsection{Centralized vs. Decentralized Actors}

As discussed in Section~\ref{sec_agent_design}, this paper introduces a novel action design termed \textit{adjust all phases}. This design allows for the modification of all phase durations within a cycle, either by increasing or decreasing their current lengths. Inspired by the action design in \textit{adjust single phase} \cite{liang2019deep, du2022safelight}, this approach enumerates all possible combinations of phase changes to form a vast action space. Subsequently, a centralized actor predicts the action from this expanded action space. However, as the number of phases at an intersection grows, the dimensionality of the resulting action space increases exponentially. To address this issue, we propose the use of decentralized actors, where each actor adjusts the duration of its corresponding traffic phase. Consequently, the policy representation is illustrated in Eq.~\eqref{eq_decentralized_policy}:

\begin{equation} \label{eq_decentralized_policy}
    \bm{\pi}_{\mathbf{\theta}}(\mathbf{a}|\mathbf{s}) 
    = 
    \left[ \pi_{\theta_{1}}(a^{1}|\mathbf{s}), \pi_{\theta_{2}}(a^{2}|\mathbf{s}) \cdots, \pi_{\theta_{N}}(a^{N}|\mathbf{s}) \right],
\end{equation}
where $a^{i}$ denotes the action for the $i$-th phase, and $\pi_{\theta_{i}}$, the actor for the $i$-th phase, is abbreviated as $\pi^{i}$ subsequently. Here, $\theta_{i}$ represents the parameters of the actor for the $i$-th phase. Fig.~\ref{fig_centralized_decentralized} illustrates the difference between the centralized actor and the decentralized actors in a two-phase signal light system, where the duration of $P_{1}$ increases by $5$~s and the duration of $P_{2}$ decreases by $5$~s. In this scenario, the value of each phase's adjustment is selected from $\{-5, 0, 5\}$. Decentralized actors enable the independent selection of adjustment values for each phase. Mathematically, if an intersection has $N$ phases  with $M$ feasible adjustment values each, the action space would be $M^N$ for a centralized actor, compared to $N \times M$ for the decentralized actors.

\begin{figure*}[!ht]
  \centering
  \includegraphics[width=0.96\textwidth]{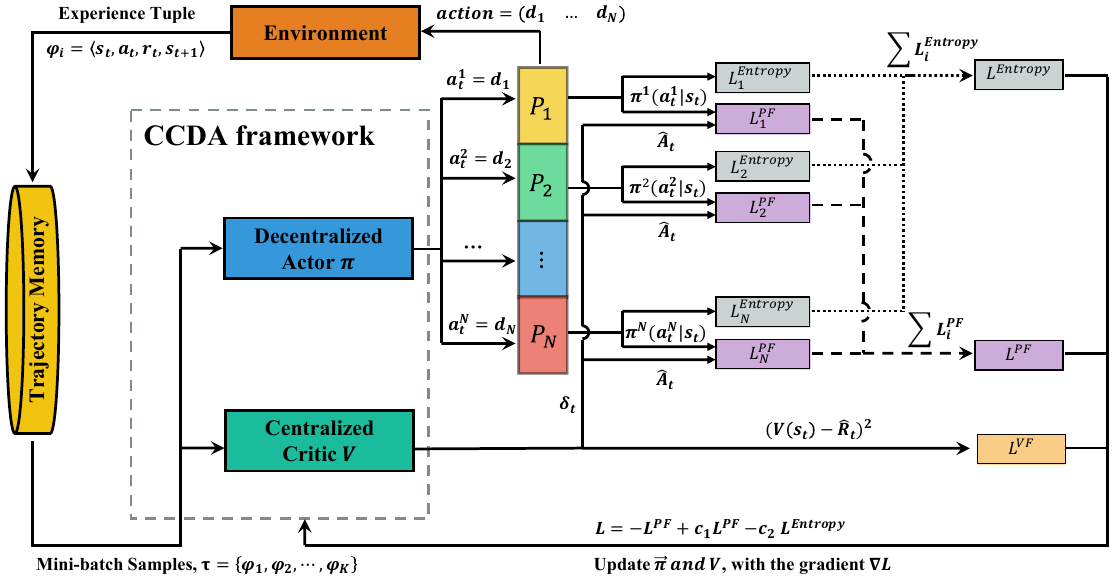}
  \caption{The policy optimization process in our Centralized Critic with Decentralized Actors (CCDA) framework.}
  \label{fig_ccda_framework}
\end{figure*}

\subsubsection{Policy Optimization for Decentralized Actors}

This section explores the method used to optimize the policy for decentralized actors within our joint traffic phase control framework. We utilize a decentralized variant of the Proximal Policy Optimization (PPO) algorithm \cite{schulman2017proximal}, specifically adapted for this traffic signal cycle control scenario. Each decentralized actor operates a neural network that predicts the value of specific actions.

Fig.~\ref{fig_ccda_framework} illustrates the policy optimization process in our CCDA framework. Initially, the CCDA model collects a mini-batch of sequential samples $\tau = \{\varphi_{1}, \cdots, \varphi_{K}\}$ from the trajectory memory. Each sample, $\varphi_{t}$, includes a tuple consisting of the state $\mathbf{s}_{t}$, action $\mathbf{a}_{t}$, reward $r_{t}$, and the subsequent state $\mathbf{s}_{t+1}$. While all tuples in the mini-batch are continuous, the selection of the first tuple can be random. Following this, the decentralized actor model predicts actions based on the observed states. Specifically, the agent determines the action for each phase independently. For example, the action for the $i$-th phase $P_{i}$ is $a_{t}^{i} = d_{i}$, where $d_i > 0$ signifies an increase in the duration of the $i$-th phase at time $t$. The policy gradient for each action is then computed. Unlike the traditional actor-critic algorithm, which uses a direct policy gradient \cite{mnih2016asynchronous}, this study employs importance sampling to estimate the expected value of samples drawn from an older policy while enhancing the new policy. The objective function for the $i$-th phase $P_{i}$ is formulated as follows:

\begin{equation} \label{policy_loss_i}
    \mathit{L}^{PF}_{i}
    = 
    \hat{\mathbb{E}} \left[ \min( \rho^{i}_{t} \hat{A_{t}}, \text{clip}( \rho^{i}_{t}, 1-\epsilon, 1+\epsilon) \hat{A_{t}})\right],
\end{equation}
where $\hat{\mathbb{E}}$ represents the empirical average over a mini-batch of samples, and $\hat{A}_{t}$ is the estimated advantage function at time step $t$. The function $\text{clip}$ ensures that the learning updates remain small and controlled. It modifies the standard policy gradient objective by introducing a clipping mechanism that restricts the ratio of the probabilities from the current policy to the old policy. Additionally, $\rho^{i}_{t}$ is the probability ratio between the old policy $\pi^{i}_{old}$ and the current policy $\pi^{i}$, defined as:
\begin{equation}
    \rho^{i}_{t}
    = 
    \frac{\pi^{i}(a_{t}^{i}|\mathbf{s}_{t})}{\pi^{i}_{old}(a_{t}^{i}|\mathbf{s}_{t})}.
\end{equation}

In this study, the generalized advantage function (GAE) \cite{schulman2015high} is used to calculate $\hat{A}_{t}$, as detailed in Eq.~\eqref{eq_advantage}: 

\begin{equation} \label{eq_advantage}
    \hat{A}_{t} 
    = 
    \delta_{t} + (\gamma \lambda) \delta_{t+1} + \cdots + (\gamma \lambda)^{K-t+1} \delta_{K-1},  
\end{equation}
where $\delta_{t} = r_{t} + \gamma V(\mathbf{s}_{t+1}) - V(\mathbf{s}_{t})$, $\lambda \in [0, 1]$ is the GAE parameter, and $K$ is the size of mini-batch samples. Here $V$ denotes the centralized critic, which evaluates the overall scenario. After calculating the policy gradient for each action $\mathit{L}^{PF}_{i}$, we propose to compute the policy gradient of the entire model $\mathit{L}^{PF}$ by summing all $\mathit{L}^{PF}_{i}$ as follows: 

\begin{equation} \label{policy_loss}
    \mathit{L}^{PF} = \sum_{i}^{N}{\mathit{L}^{PF}_{i}}. 
\end{equation}

To enhance the exploration capabilities of the agent, an entropy term is added to the objective function. This encourages the agent to explore a wider range of actions, increasing the randomness of its decisions and preventing premature convergence on suboptimal policies \cite{yuan2023renyi}. For the actor of phase $i$, the entropy loss $\mathit{L}^{Entropy}_{i}$ is computed based on the policy $\pi^{i}(\cdot|\mathbf{s})$ as follows:
\begin{equation} \label{entropy_loss_i}
    \mathit{L}^{Entropy}_{i}
    = 
    \sum_{a_{i} \in \mathcal{A}} \pi^{i}(a_{i}|\mathbf{s}_{t}) log \left( \pi^{i}(a_{i}|\mathbf{s}_{t}) \right). 
\end{equation}

In a similar manner to calculating $\mathit{L}^{PF}$, we aggregate the entropy losses from all phases to determine the total entropy loss $\mathit{L}^{Entropy}$ for the entire model:
\begin{equation} \label{entropy_loss}
    \mathit{L}^{Entropy}
    = 
    \sum_{i}^{N}{\mathit{L}^{Entropy}_{i}}. 
\end{equation}

\subsubsection{Centralized Critic for Cooperative Traffic Phase Management}

In the preceding section, we computed the policy loss and entropy loss for each traffic phase actor and subsequently aggregated them. Initially, the actions of the traffic phase actors are independent. To enhance collaboration among these actors, we introduced a centralized critic. As illustrated in Fig.~\ref{fig_ccda_framework}, the centralized critic evaluates the current scene, independent of the number of traffic phase actors.

To update the centralized critic, we aim to minimize the discrepancy between the estimated and actual values. Following the methodology in \cite{arulkumaran2017deep, mnih2016asynchronous, moerland2023model}, this study employs the objective function $\mathit{L}^{VF}$, defined as the squared-error loss between the value function $V(t)$ and the cumulative return $\hat{R}_{t}$:

\begin{equation} \label{value_loss}
    \mathit{L}^{VF}
    = 
    \hat{\mathbb{E}} \left[ (V(s_{t}) - \hat{R}_{t})^{2} \right], 
\end{equation}
where 
\begin{equation}
	\hat{R}_{t} = \sum_{k=0}^{\infty}{\gamma^{k} r_{t+k}},
\end{equation}
with $\gamma \in [0,1)$ representing the discount factor that balances immediate and future rewards, and $r_{t}$ being the reward calculated according to Eq.~\eqref{eq_reward}.

Upon calculating the value loss for the centralized critic, the final objective function is expressed as shown in Eq.~\eqref{final_object_function}:

\begin{equation} \label{final_object_function}
    \mathit{L} = - \mathit{L}^{PF} + c_{1} \mathit{L}^{VF} - c_{2}  \mathit{L}^{Entropy},  
\end{equation}
where $c_{1}$ and $c_{2}$ represent the coefficients for the critic loss and the entropy bonus, respectively. The final objective function, $\mathit{L}$, comprises three components: the policy loss from decentralized actors, which aids in decision-making; the entropy loss from each phase actor, which encourages exploration; and the value loss from the centralized critic, which assesses scenarios to facilitate collaboration.

\begin{figure}[!ht]
  \centering
  \includegraphics[width=0.6\linewidth]{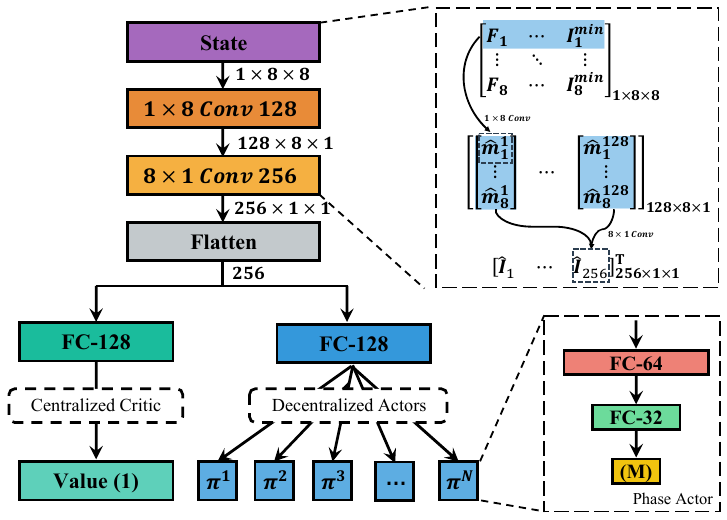}
  \caption{The architecture of the neural network used in AAP (CCDA).}
  \label{fig_model_architecture}
\end{figure}

\subsection{Model Structure for AAP (CCDA)} \label{model_architecture}

Fig.~\ref{fig_model_architecture} depicts the detailed neural network architecture proposed in this work. The network receives an input state $\mathbf{s} \in \mathbb{R}^{8 \times 8}$, which comprises the features of $8$ movements as described in Section~\ref{sec_agent_design}. This state $\mathbf{s}$ is processed by a feature extractor neural network, which is shared between the decentralized actors and the centralized critic. To extract relevant information from the intersections, the network employs two convolutional layers. The first layer is a 2D convolution with $128$ filters of size $1 \times 8$, designed to gather information for each movement, resulting in $\widehat{\mathbf{m}}_{i} = \left[\widehat{m}_{i}^{1}, \widehat{m}_{i}^{2}, \ldots, \widehat{m}_{i}^{128} \right]$. The second layer is a 2D convolution with $256$ filters of size $8 \times 1$, which extracts information specific to the intersection, denoted by $\hat{\bm{I}} = \left[\hat{I}_{1}, \hat{I}_{2}, \ldots, \hat{I}_{256} \right]$.

Following the feature extraction, $\hat{\bm{I}}$ captures the comprehensive information of the intersection, which is utilized by the phase actors to make decisions and by the centralized critic to assess the current environment. The centralized critic model, located in the lower left corner of Fig.~\ref{fig_model_architecture}, consists of two fully-connected layers. This model outputs a scalar value that evaluates the actions taken by all phase agents. In the lower right corner of Fig.~\ref{fig_model_architecture}, the decentralized actors for each traffic phase are presented. Each phase actor comprises a three-layer fully connected network that outputs an array of size $M$, indicating that each phase can select from $M$ possible values to adjust the phase duration. Algorithm~\ref{alg_ccda_rl_tsc} details the procedure that policy optimization process in our CCDA framework.

\begin{algorithm}[!ht]
    \setstretch{1.2}
    \caption{AAP (CCDA) Training Workflow}
    \label{alg_ccda_rl_tsc}
    \KwInput{Number of phases $N$, initial phase duration $[d_{1}, \cdots, d_{N}]$, maximum simulation time $T$, intervention frequency $\Delta t$, mini-batch size $K$, coefficient of loss function $c_{1}$, $c_{2}$.}
    \For{$i=1,2,\cdots$}{
        Initialize an empty trajectory buffer $\mathcal{B}$ \;
        \While{$t < T$}{
            \For{$j=1,2, \cdots N$}{
                $a_{t}^{j} \sim \pi^{j}(\cdot|\mathbf{s}_{t})$ \;
                $d_{j} \gets d_{j} + a_{t}^{j}$ \; % \Comment*[r]{Adjust duration for phase $P_{j}$}
            }
            $\mathbf{a}_{t} \gets (a_{t}^{1}, \cdots, a_{t}^{N})$ \; 
            % calculate delta time
            $C \gets \sum_{i=1}^{N}{d_{i}}$ % \Comment*[r]{Calculate cycle length}
            $\Delta t^{\prime} \gets \left\lceil \frac{\Delta t}{C} \right\rceil \times C$ \; 
            Wait $\Delta t^{\prime}$ seconds until the end of cycle \;
            Observe next state $\mathbf{s}_{t+1} \sim p(\cdot | \mathbf{s}_{t}, \mathbf{a}_{t})$ \;
            Append $(\mathbf{s}_{t}, \mathbf{a}_{t}, r(\mathbf{s}_{t}, \mathbf{a}_{t}), \mathbf{s}_{t+1})$ to $\mathcal{B}$ \;
            Update simulation time $t \gets t + \Delta t^{\prime}$ % \Comment*[r]{Update simulation time}
            
            \If{$|\mathcal{B}| \geq K$}{
                Collect a mini-batch $\tau$ from $\mathcal{B}$ \;
                \For{$j=1,2, \cdots N$}{
                    Compute $\mathit{L}^{PF}_{j}$ according to Eq.~\eqref{policy_loss_i} \; % policy loss
                    Compute $\mathit{L}^{Entropy}_{j}$ according to Eq.~\eqref{entropy_loss_i} % entropy loss
                }
                $\mathit{L}^{PF} \gets \sum_{j=1}^{N}{\mathit{L}^{PF}_{j}}$ \; % \Comment*[r]{Policy loss}
                $\mathit{L}^{Entropy} \gets \sum_{j=1}^{N}{\mathit{L}^{Entropy}_{j}}$ \; % \Comment*[r]{Entropy loss}
                $\mathit{L}^{VF} \gets \frac{1}{K} \sum_{k=1}^{K} \left( V_{\theta}(s_{k}) - \hat{R}_{k} \right)^{2}$ \; % \Comment*[r]{Value loss}
                $\mathit{L} \gets - \mathit{L}^{PF} + c_{1} \mathit{L}^{VF} - c_{2}  \mathit{L}^{Entropy}$ \;
                Update $\bm{\pi}$ and $V$ with $\nabla \mathit{L}$ via Adam \;
                Clear $\mathcal{B}$ \;
            }
        }
    }
\end{algorithm}

% %%%%%%%%%%%
% Experiments
% %%%%%%%%%%%
\section{Experiment Setup} \label{sec_experiment}

\subsection{Experiment Setting}

The experiments were conducted using Simulation of Urban MObility (SUMO) version 1.10 \cite{alvarezlopezMicroscopicTrafficSimulation2018}, a comprehensive open-source traffic simulation platform. Traffic signal management and data collection, including occupancy and queue lengths, were handled using SUMO's Traffic Control Interface (TraCI). However, the scope of data collection was limited to a 150-meter radius from the intersections, reflecting the range limitation of cameras in the simulated environment. As a result, the queue lengths reported in this study are capped at 150 meters, irrespective of potential longer queues.

The simulation infrastructure consisted of a high-performance computing system with an AMD EPYC 7763 CPU, 256 GB of RAM, and an NVIDIA RTX 3090Ti GPU, all running on Ubuntu 20.04 LTS. Within the simulation, right-turn movements were permitted at all times, and transitions from a green to a red signal were intermediated by a 3-second yellow light. Following established research \cite{wei2019survey}, the control agent was programmed to alter the traffic signals at 5-second intervals, indicating a high interaction frequency. These interactions are visualized in Fig.~\ref{fig_choose_next_phase} and Fig.~\ref{fig_next_or_not}, with \textit{adjust all phases} actions occurring at $\{-6, -3, 0, 3, 6\}$ second intervals. A sensitivity analysis under different magnitudes of change was also conducted in Section~\ref{sec_sensitivity_analysis}.

To mirror real-world urban traffic conditions, several critical parameters were incorporated. The maximum speed limit on simulated roads was set to $13.9$ m/s, approximately $50$ km/h. A minimum vehicle gap of $2.5$ m was enforced, and vehicle speed distribution was modeled as a Gaussian distribution with a mean of $10$ m/s and a variance of $3$. These parameters were selected to emulate the behavior of vehicles under typical urban traffic scenarios. The detailed network parameters used for this research are included in Table~\ref{tab_parameters}.

\begin{table}[!ht]
    \centering
    \caption{Parameters in the AAP (CCDA) framework for TSC.}
    \label{tab_parameters}
    \begin{tabular}{lc}
    \hline
    Parameter & Value \\ \hline
    Learning rate $\eta$ & $0.0001$ \\
    Trajectory memory size $M$ & $3000$ \\
    Mini-batch size $K$ & $256$ \\
    Clipping range $\epsilon$ in Eq.~\eqref{policy_loss_i} & $0.2$ \\
    Discount factor $\gamma$ in Eq.~\eqref{eq_advantage} & $0.99$ \\
    GAE parameter $\lambda$ in Eq.~\eqref{eq_advantage} & $0.95$ \\
    Value function coefficient $c_{1}$ in Eq.~\eqref{final_object_function}  & $0.9$ \\ 
    Entropy coefficient $c_{2}$ in Eq.~\eqref{final_object_function} & $0.01$ \\ \hline
    \end{tabular}
\end{table}

\subsection{Evaluation Metrics}

To effectively assess the proposed framework, this study employs two key metrics: measurement of intersection performance and measurement of signal system stability.

\textbf{Efficiency}: 
Recalling that the average queue length $q_{i}^{t}$ is calculated based on the endpoint of the last standing vehicle, we propose the following efficiency metric:
\begin{equation} \label{eq:m_efficiency}
    m_{q} = \frac{\sum_{t=1}^{T} \sum_{i=1}^{M}{q_{i}^{t}}}{T \times M},
\end{equation}
where $M$ represents the total number of lanes in the scenario and $T$ represents the entire simulation duration. 

\textbf{Steadiness}: 
In real-world applications, the stability of the traffic signal plans is crucial as it ensures that the signal plan does not change dramatically over a short period. We utilize the second-order difference to quantify stability, as it effectively captures the fluctuation in a time sequence's alterations \cite{bouadi2022stability}. This metric not only measures the variability of signal changes but also serves as an index of overall system stability, where fewer changes denote higher stability. For the TSC system, we define the second-order difference for phase $n$ between periods $i$ and $i+2$ as:

\begin{equation}
\begin{split}
    \Delta^2 d_{n}^{i} & = (d_{n}^{i+2} - d_{n}^{i+1}) - (d_{n}^{i+1} - d_{n}^{i}) \\
    & = d_{n}^{i+2} - 2d_{n}^{i+1} + d_{n}^{i},
\end{split}
\end{equation}
where $N$ is the total number of phases within a traffic signal cycle, $K$ is the count of periods under observation, and $d_{n}^{i}$ denotes the duration of the $n$-th phase in the $i$-th period, with $n \in \{1, 2, \ldots, N\}$ and $i \in \{1, 2, \ldots, K\}$. To construct a stability metric using second-order differences, we aggregate the absolute values of these differences over all applicable periods and phases:

\begin{equation}
    \Delta^2_{total} = \sum_{n=1}^{N} \sum_{i=1}^{K-2} |\Delta^2 d_{n}^{i}|.
\end{equation}

The stability indicator $m_{s}$, based on the second-order difference, is then expressed as:

\begin{equation} \label{eq:m_steadiness}
    m_{s} = \frac{\Delta^2_{total}}{D_{total}},
\end{equation}
where $D_{total}$ is the sum of the durations of all phases for all periods:

\begin{equation}
    D_{total} = \sum_{n=1}^{N} \sum_{i=1}^{K} d_{n}^{i}.
\end{equation}

This proportion offers a normalized metric of the second-order differences in relation to the total duration of all phases across all periods. A lower $m_{s}$ indicates greater stability in the traffic signal timing, with minimal variation in the phase durations. Conversely, a higher $m_{s}$ indicates less stability, with more pronounced changes in phase durations.

\subsection{Dataset}

% Synthetic data
\subsubsection{Synthetic Data}

\begin{figure}[!ht]
  \centering   
  % Four Way Intersection
  \subfloat[]{\includegraphics[width=0.4\textwidth]{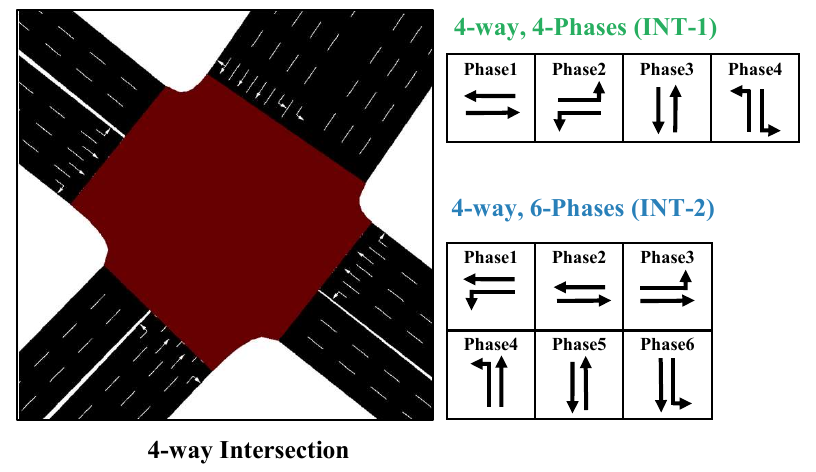}
  \label{fig:syn_int12}}

  \hfill
  
  % Three Way Intersection
  \subfloat[]{\includegraphics[width=0.4\textwidth]{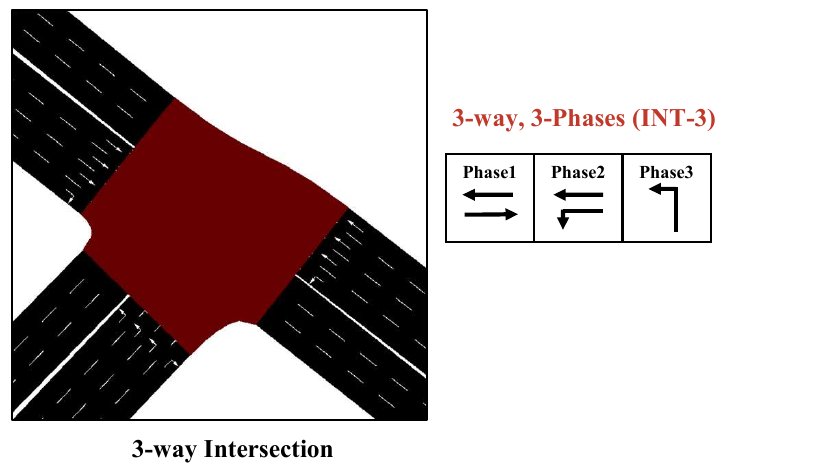}
  \label{fig:syn_int3}}

  \caption{Topologies and phases for three synthetic intersections. (a) The two synthetic 4-way intersections with 4 and 6 phases, respectively. (b) The synthetic 3-way intersection with 3 phases.}
    \label{fig:syn_network}
\end{figure}

In the first part of our experiment, synthetic data with various intersections and routes are employed. Specifically, three intersections of different numbers of phases and approaching roads (e.g., 3-way and 4-way intersections) are constructed and used for experiments. As depicted in Fig.~\ref{fig:syn_network}, the topologies and phases of the three intersections are described as follows: \textit{INT-1}, a 4-way intersection with 4 phases; \textit{INT-2}, a 4-way intersection with 6 phases; and \textit{INT-3}, the 3-way intersection scenario with 3 phases, respectively. 

For each intersection, two configurations of traffic flows are generated: a steady traffic flow (\textit{route1-X}) and a complex traffic flow (\textit{route2-X}). A summary of the traffic configurations for the six synthetic scenarios is provided in Table~\ref{tab_config_synthetic}. Each scenario lasts for a duration of two hours, with vehicle arrivals determined using a Poisson distribution based on the specified arrival rates.

\begin{table}[!ht]
    \centering
    \caption{Configurations for synthetic traffic data.}
    \label{tab_config_synthetic}
        \begin{tabular}{ccccccc}
        \hline
        \multirow{2}{*}{Network} & \multirow{2}{*}{Route} & \multirow{2}{*}{Records} & \multicolumn{4}{c}{Arrival Rate (vehicles/s)} \\ \cmidrule{4-7} % cline
         &  &  & Mean & Std & Min & Max \\ \hline
        \multirow{2}{*}{INT-1} & Route1-1 & $6784$ & $0.942$ & $0.061$ & $0.733$ & $1.133$ \\
         & Route2-1 & $6986$ & $0.970$ & $0.092$ & $0.811$ & $1.052$ \\ \hline
        \multirow{2}{*}{INT-2} & Route1-2 & $6774$ & $0.941$ & $0.067$ & $0.828$ & $1.282$ \\
         & Route2-2 & $6755$ & $0.938$ & $0.102$ & $0.871$ & $1.272$ \\ \hline
        \multirow{2}{*}{INT-3} & Route1-3 & $4786$ & $0.665$ & $0.052$ & $0.563$ & $0.709$ \\
         & Route2-3 & $5530$ & $0.768$ & $0.098$ & $0.716$ & $0.801$ \\ \hline
        \end{tabular}
\end{table}

\begin{figure}[!ht]
  \centering   
  \subfloat[]{\includegraphics[width=0.4\textwidth]{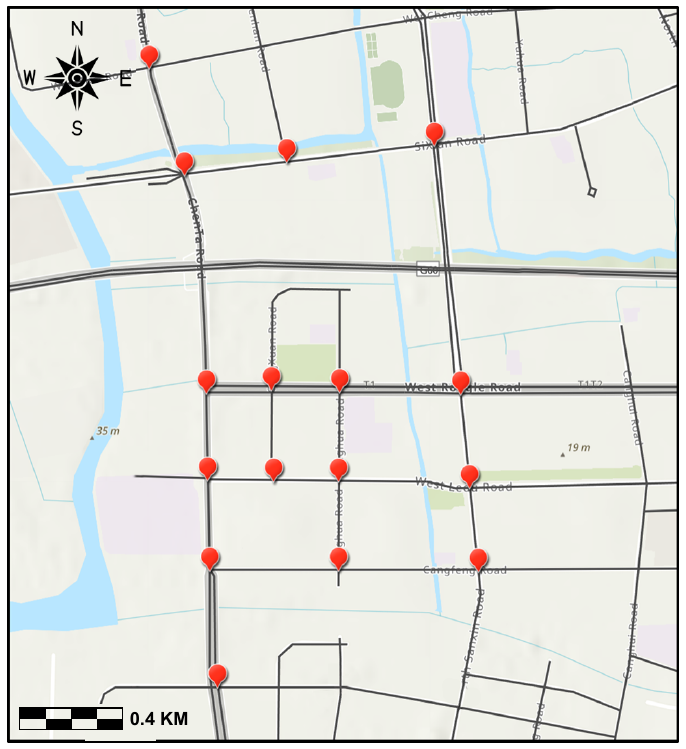}%
  \label{fig_real_world_map}}

  \hfill
  
  \subfloat[]{\includegraphics[width=0.4\textwidth]{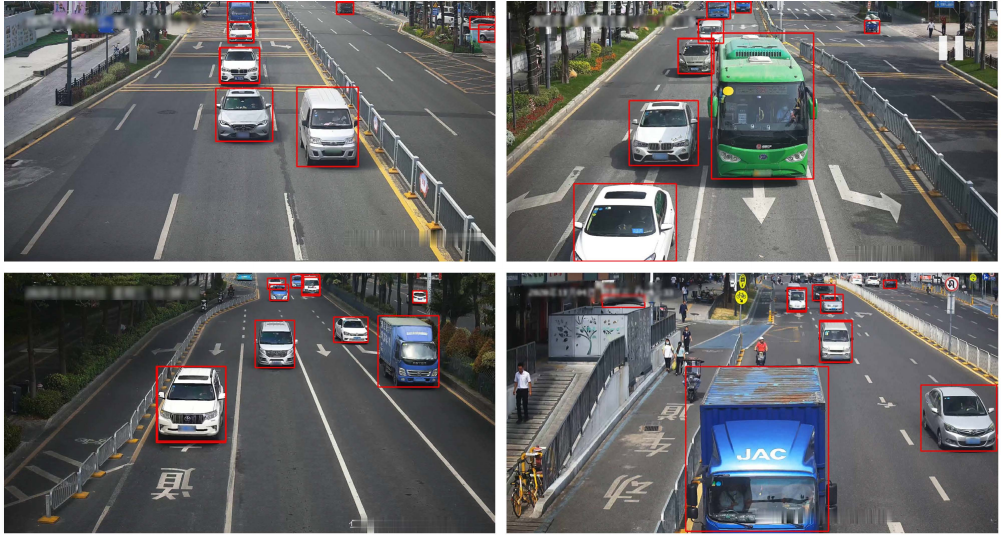}%
  \label{fig_vehicle_cctv}}
  \caption{Traffic Network at Chenta Road, Songjiang District, Shanghai, China: (a) SUMO Simulation Network; (b) Surveillance Camera Video from Four Directions at a Single Intersection.}
  \label{fig_real_world_network}
\end{figure}

\begin{table}[!ht]
    \centering
    \caption{Configurations for real-world traffic data.} \label{tab_config_real_world}
    \begin{tabular}{cccccc}
    \hline
    \multirow{2}{*}{Network} & \multirow{2}{*}{Records} & \multicolumn{4}{c}{Arrival Rate (vehicles/s)} \\ \cmidrule{3-6} 
     &  & Mean & Std & Min & Max \\ \hline
     3-way & $221,945$ & $0.367$ & $0.239$ & $0.014$ & $0.695$ \\
     4-way & $377,786$ & $0.486$ & $0.312$ & $0.021$ & $0.946$ \\  \hline
    \end{tabular}
\end{table}

% Real-world data
\subsubsection{Real-world Data}

As illustrated in Fig.~\ref{fig_real_world_map}, a traffic network located near the Chenta Road in the Songjiang district of Shanghai, China is selected as the study area. This region is characterized by a high density of buildings and businesses, which contributes to significant traffic congestion. The traffic network within this area includes $16$ intersections, comprising nine 4-way intersections and seven 3-way intersections. Traffic surveillance videos from these intersections, recorded on July 30, 2021, were analyzed by YOLOv7 \cite{wang2023yolov7} to estimate the vehicle flow per minute at each intersection, as depicted in Fig.~\ref{fig_vehicle_cctv}. These estimates were then used as inputs in the SUMO simulations for our experiments. The variations in arrival rates at the 3-way and 4-way intersections are detailed in Table~\ref{tab_config_real_world}, reflecting the dynamic nature of the traffic flow in this real-world setting.

\begin{figure*}[!ht]
    \centering
    \subfloat[]{\includegraphics[width=0.48\textwidth]{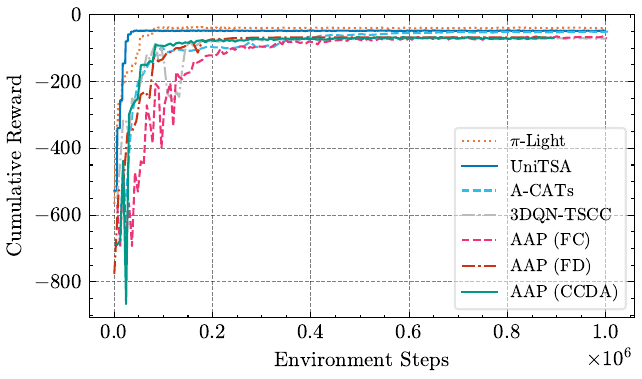}
    \label{fig_rc_None}}
    \subfloat[]{\includegraphics[width=0.48\textwidth]{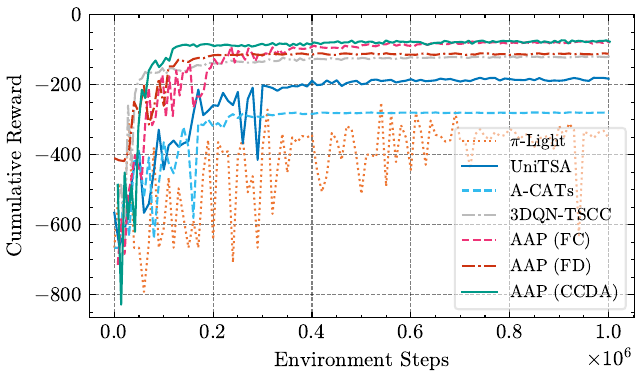}
    \label{fig_rc_60}}
    
    \hfill
    
    \subfloat[]{\includegraphics[width=0.48\textwidth]{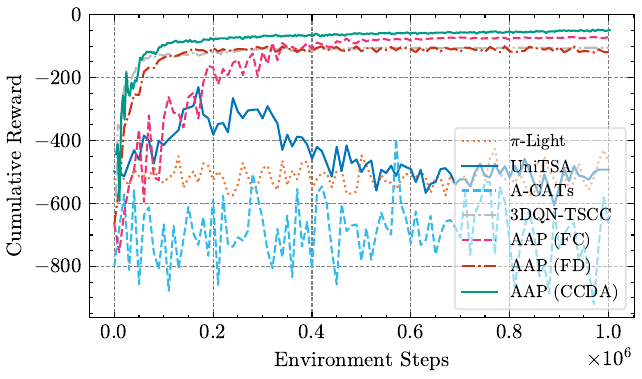}
    \label{fig_rc_120}}
    \subfloat[]{\includegraphics[width=0.48\textwidth]{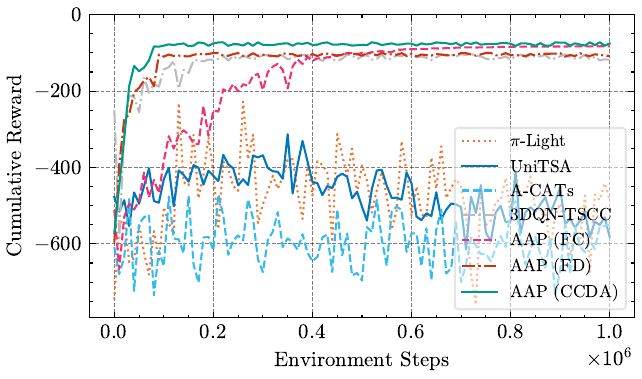}
    \label{fig_rc_300}}
    
    \caption{Cumulative reward curve during training for INT-1, Route1-1 under various $\Delta t$ between RL-based TSC methods. (a) $\Delta t=0$. (b) $\Delta t=60$. (c) $\Delta t=120$. (d) $\Delta t=300$.}
    \label{fig_reward_curves}
\end{figure*}

\begin{table*}[!ht]
    \centering
    \caption{Traffic efficiency under varying intervention frequencies with different methods on synthetic and real-world data. A lower value indicates better performance.}
    \label{tab_results_efficiency}
    \resizebox{\textwidth}{!}{%
    \begin{tabular}{lccccccccc}
    \hline
     & \multicolumn{6}{c}{\textbf{Synthetic Data}} &  & \multicolumn{2}{c}{\textbf{Real-world Data}} \\ \cline{2-7} \cline{9-10} 
    \multicolumn{1}{c}{} & \multicolumn{2}{c}{\textit{\textbf{INT-1}}} & \multicolumn{2}{c}{\textit{\textbf{INT-2}}} & \multicolumn{2}{c}{\textit{\textbf{INT-3}}} & \textit{} & \multicolumn{2}{c}{\textit{\textbf{ChenTa Road}}} \\ \cline{2-7} \cline{9-10} 
    Method & \textit{Route1-1} & \textit{Route2-1} & \textit{Route1-2} & \textit{Route2-2} & \textit{Route1-3} & \textit{Route2-3} &  & \textit{3-Way} & \textit{4-Way} \\ \hline
     & \multicolumn{9}{c}{$\Delta t = 0$} \\
    FT-30 & $16.309_{\pm 2.3}$ & $19.633_{\pm 2.1}$ & $22.247_{\pm 2.6}$ & $26.411_{\pm 3.1}$ & $6.583_{\pm 0.5}$ & $13.607_{\pm 1.8}$ &   & $2.982_{\pm 1.0}$ & $17.734_{\pm 15.2}$ \\
    FT-40 & $21.249_{\pm 2.7}$ & $24.409_{\pm 3.3}$ & $29.509_{\pm 4.2}$ & $33.214_{\pm 1.8}$ & $8.899_{\pm 1.2}$ & $14.571_{\pm 2.1}$ &   & $3.781_{\pm 0.9}$ & $22.974_{\pm 9.8}$ \\
    Webster & $13.783_{\pm 1.8}$ & $21.477_{\pm 2.7}$ & $15.434_{\pm 2.0}$ & $38.431_{\pm 3.0}$ & $4.366_{\pm 0.6}$ & $8.231_{\pm 1.1}$ &   & $1.129_{\pm 0.3}$ & $8.344_{\pm 5.7}$ \\
    $\pi$-Light & $5.042_{\pm 0.7}$ & $5.401_{\pm 0.8}$ & $3.913_{\pm 0.5}$ & $5.139_{\pm 0.7}$ & $1.327_{\pm 0.2}$ & $2.098_{\pm 0.1}$ &   & $0.484_{\pm 0.2}$ & $3.716_{\pm 1.4}$ \\
    UniTSA & $5.351_{\pm 0.8}$ & $5.728_{\pm 1.4}$ & $4.688_{\pm 0.6}$ & $6.652_{\pm 0.4}$ & $1.445_{\pm 0.1}$ & $2.573_{\pm 0.2}$ &   & $0.654_{\pm 0.2}$ & $4.128_{\pm 2.5}$ \\
    A-CATs & $6.308_{\pm 0.7}$ & $6.266_{\pm 0.8}$ & $6.994_{\pm 0.5}$ & $7.089_{\pm 0.8}$ & $1.534_{\pm 0.2}$ & $2.932_{\pm 0.4}$ &   & $0.783_{\pm 0.2}$ & $4.983_{\pm 2.7}$ \\
    3DQN-TSCC & $8.915_{\pm 0.5}$ & $8.764_{\pm 0.1}$ & $11.828_{\pm 1.6}$ & $12.106_{\pm 1.4}$ & $2.127_{\pm 0.3}$ & $3.685_{\pm 0.5}$ &   & $0.826_{\pm 0.6}$ & $5.893_{\pm 4.5}$ \\
    AAP (FC) & $9.067_{\pm 0.9}$ & $8.539_{\pm 0.9}$ & $13.887_{\pm 1.7}$ & $11.347_{\pm 0.8}$ & $1.891_{\pm 0.8}$ & $3.505_{\pm 0.2}$ &   & $0.656_{\pm 0.6}$ & $5.612_{\pm 4.7}$ \\
    AAP (FD) & $9.355_{\pm 1.4}$ & $9.276_{\pm 0.7}$ & $12.139_{\pm 1.3}$ & $13.004_{\pm 1.7}$ & $2.296_{\pm 0.5}$ & $3.852_{\pm 0.4}$ &   & $0.991_{\pm 0.6}$ & $5.956_{\pm 3.5}$ \\
    AAP (CCDA) & $8.693_{\pm 0.8}$ & $9.001_{\pm 0.9}$ & $9.576_{\pm 0.6}$ & $9.797_{\pm 1.0}$ & $1.803_{\pm 0.2}$ & $3.383_{\pm 0.2}$ &   & $0.678_{\pm 0.5}$ & $5.393_{\pm 4.3}$ \\ \hline
     & \multicolumn{9}{c}{$\Delta t = 60$} \\
    3DQN-TSCC & $9.477_{\pm 0.6}$ & $9.029_{\pm 0.5}$ & $11.543_{\pm 1.4}$ & $11.856_{\pm 1.7}$ & $2.445_{\pm 0.3}$ & $4.614_{\pm 0.7}$ &   & $0.843_{\pm 0.6}$ & $6.292_{\pm 5.9}$ \\
    AAP (FC) & $9.616_{\pm 1.1}$ & $9.827_{\pm 1.3}$ & $8.143_{\pm 0.7}$ & $12.386_{\pm 1.4}$ & $2.637_{\pm 0.5}$ & $4.095_{\pm 0.4}$ &   & $0.712_{\pm 0.6}$ & $5.737_{\pm 4.6}$ \\
    AAP (FD) & $9.248_{\pm 1.1}$ & $9.372_{\pm 1.0}$ & $9.019_{\pm 0.5}$ & $13.104_{\pm 0.8}$ & $2.896_{\pm 0.9}$ & $7.852_{\pm 0.5}$ &   & $0.974_{\pm 0.9}$ & $7.545_{\pm 7.3}$ \\
    AAP (CCDA) & $9.389_{\pm 1.2}$ & $9.442_{\pm 1.2}$ & $10.405_{\pm 0.5}$ & $9.361_{\pm 1.2}$ & $2.265_{\pm 0.3}$ & $6.229_{\pm 0.2}$ &   & $0.694_{\pm 0.3}$ & $5.928_{\pm 0.6}$ \\ \hline
     & \multicolumn{9}{c}{$\Delta t = 120$} \\
    3DQN-TSCC & $13.483_{\pm 1.8}$ & $13.778_{\pm 1.3}$ & $12.296_{\pm 1.2}$ & $12.524_{\pm 1.1}$ & $2.749_{\pm 0.2}$ & $3.838_{\pm 0.4}$ &   & $2.922_{\pm 3.9}$ & $8.851_{\pm 7.5}$ \\
    AAP (FC) & $10.446_{\pm 1.5}$ & $8.695_{\pm 1.1}$ & $59.009_{\pm 3.2}$ & $10.274_{\pm 1.3}$ & $8.58_{\pm 2.1}$ & $4.191_{\pm 1.1}$ &   & $1.225_{\pm 0.7}$ & $5.756_{\pm 4.1}$ \\
    AAP (FD) & $14.575_{\pm 2.1}$ & $12.956_{\pm 1.3}$ & $17.729_{\pm 2.6}$ & $11.482_{\pm 1.7}$ & $3.196_{\pm 0.4}$ & $3.825_{\pm 0.6}$ &   & $1.527_{\pm 0.4}$ & $6.261_{\pm 5.8}$ \\
    AAP (CCDA) & $11.021_{\pm 1.2}$ & $12.19_{\pm 1.6}$ & $9.905_{\pm 0.7}$ & $8.914_{\pm 1.2}$ & $2.344_{\pm 0.3}$ & $4.486_{\pm 0.4}$ &   & $1.162_{\pm 0.6}$ & $5.625_{\pm 4.1}$ \\ \hline
     & \multicolumn{9}{c}{$\Delta t = 300$} \\
    3DQN-TSCC & $14.511_{\pm 1.8}$ & $13.739_{\pm 1.3}$ & $15.986_{\pm 2.2}$ & $16.842_{\pm 1.9}$ & $3.286_{\pm 0.8}$ & $4.570_{\pm 0.5}$ &   & $4.321_{\pm 3.7}$ & $13.536_{\pm 11.8}$ \\
    AAP (FC) & $8.355_{\pm 2.2}$ & $8.596_{\pm 1.8}$ & $25.709_{\pm 6.8}$ & $33.004_{\pm 4.8}$ & $2.296_{\pm 0.3}$ & $3.852_{\pm 0.4}$ &   & $1.794_{\pm 0.6}$ & $8.432_{\pm 7.7}$ \\
    AAP (FD) & $18.051_{\pm 0.7}$ & $15.734_{\pm 1.4}$ & $19.119_{\pm 2.4}$ & $13.461_{\pm 1.5}$ & $3.901_{\pm 0.2}$ & $4.421_{\pm 0.7}$ &   & $2.294_{\pm 0.6}$ & $9.168_{\pm 6.7}$ \\
    AAP (CCDA) & $7.911_{\pm 0.7}$ & $7.507_{\pm 0.4}$ & $8.127_{\pm 0.6}$ & $8.714_{\pm 0.6}$ & $2.441_{\pm 0.3}$ & $3.375_{\pm 0.2}$ &   & $1.813_{\pm 0.6}$ & $7.679_{\pm 6.8}$ \\ \hline
    \end{tabular}}
\end{table*}

\subsection{Benchmarking Methods}

To validate the performance of the proposed framework, we conducted comparative analyses with established baseline approaches, including both rule-based TSC methods and RL-based methods with diverse agent architectures.

\begin{itemize}
    \item \textbf{Fixed-time Control (FT)}. The fixed-time control strategy, which utilizes predetermined cycle lengths and phase durations, is widely adopted in scenarios with consistent traffic patterns \cite{miller1963settings}. For our experiments, we set the phase durations at 30 seconds (FT-30) and 40 seconds (FT-40) respectively.
    \item \textbf{Webster} \cite{kouvelas2011hybrid}. This method determines cycle durations and phase splits based on traffic volumes. For equitable comparison, we recalculated the optimal cycle length and phase durations every ten minutes using real-time traffic data with the Webster Method.
    \item \textbf{RL-based TSC}. RL-based TSC methods have shown promising results with various action selection strategies. As benchmarks, we selected the most recent methods corresponding to four distinct action designs: $\pi$-Light \cite{gu2024pi} employing the \textit{choose next phase} strategy, UniTSA \cite{wang2023unitsa} with the \textit{next or not} approach, A-CATs \cite{aslani2017adaptive} utilizing the \textit{set current phase duration} method, and 3DQN-TSCC \cite{liang2019deep} which adjusts a \textit{single phase}.
\end{itemize}

Beyond the aforementioned models, we introduce two variants of our approach that solely implement the novel action design \textit{adjust all phases} (AAP), independent of the CCDA framework.

\begin{itemize}
    \item \textbf{AAP (FC)}. This variant adopts a fully centralized (FC) scheme, where both the actor and the critic of the RL algorithm are centralized. The action space includes all potential variations in phase durations. However, the action space grows exponentially with the number of phases, potentially causing scalability issues.
    \item \textbf{AAP (FD)}. The fully decentralized (FD) model employs decentralized actors and critics. While it also leverages the \textit{adjust all phases} action design, it reduces the action space through distributed control among multiple agents. Each intersection is managed by an autonomous actor with a dedicated critic, facilitating a scalable framework. Nonetheless, the decentralized critics may compromise the coordination between intersections, possibly diminishing the efficiency of traffic phase adjustments.
    \item \textbf{AAP (CCDA)}.This approach, which is central to our paper, combines decentralized actors with a centralized critic. It allows independent action selection for each phase by the actors, while the centralized critic assesses the overall traffic network state. The aim is to merge the scalability of decentralized actors with the coordinated decision-making enabled by a centralized critic.
\end{itemize}

% %%%%%%%
% Results
% %%%%%%%
\begin{table}[!ht]
    \caption{Comparison of action spaces for different traffic control methods across various environments.}
    \label{tab_action_size}
    \centering
    \begin{tabular}{lccc}
    \hline
     & INT-1 & INT-2 & INT-3 \\ \hline
    3DQN-TSCC & $9$ & $13$ & $7$ \\
    AAP (FC) & $625$ & $15625$ & $125$ \\
    AAP (FD) \& AAP (CCDA) & $20$ & $30$ & $15$ \\ \hline
    \end{tabular}
\end{table}

\section{Results and Discussion} \label{sec_results}

In this section, we present the performance of AAP (CCDA) as compared to various benchmarking methods on both synthetic and real-world datasets under various intervention frequencies. We also examine and compare the traffic signal control policies learned by CCDA and the benchmark methods. Additionally, we perform a sensitivity analysis on the intervention frequencies to further validate the effectiveness of CCDA.

\begin{figure}[!ht]
  \centering
  \includegraphics[width=0.5\linewidth]{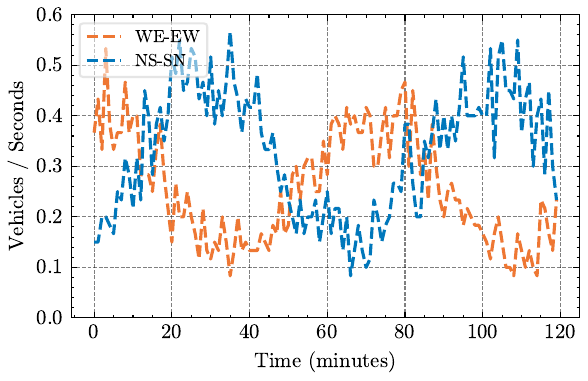}
  \caption{Traffic Flow Fluctuations at INT-1, Route2-1.}
  \label{fig_traffic_flow}
\end{figure}

\subsection{Comparison of Efficiency}

Fig.~\ref{fig_reward_curves} shows the reward curves of different methods under various values of $\Delta t$ for INT-1, Route1-1. Table~\ref{tab_results_efficiency} compares the performance of different methods under different $\Delta t$ values in terms of traffic efficiency, measured by the average queue length. As shown in Fig.~\ref{fig_rc_None}, when $\Delta t=0$, which means there is no limit on the frequency of actions, $\pi$-Light based on the \textit{choose next phase} strategy and UniTSA based on the \textit{next or not} approach achieved higher cumulative rewards. However, as $\Delta t$ increases, as shown in Fig.~\ref{fig_rc_60}, the rewards of existing methods like $\pi$-Light, UniTSA, and A-CATs decrease significantly. When $\Delta t$ grows to $120$ and $300$, as shown in Fig.~\ref{fig_rc_120} and Fig.~\ref{fig_rc_300}, meaning the agent can only adjust the traffic signal every $2$ and $5$ minutes, these methods fail to converge. As the intervention frequency decreases, it becomes increasingly challenging and time-consuming to adjust the entire signal timing plan using these methods, making them unsuitable for rapidly changing traffic volumes. Therefore, when $\Delta t > 0$, these methods are excluded from the comparisons presented in Table~\ref{tab_results_efficiency}.

Both 3DQN-TSCC and the proposed method, AAP, can converge at lower intervention frequencies. Fig.~\ref{fig_rc_120} and Fig.~\ref{fig_rc_300} show that the proposed AAP (CCDA) approach consistently achieves higher reward values than 3DQN-TSCC across all intervention frequencies. This superior performance is attributable to our method's ability to adjust all phases in one action, unlike 3DQN-TSCC, which adjusts only one phase at a time. Additionally, while methods utilizing the AAP action like AAP (FC) exhibit reward values close to those of AAP (CCDA), their convergence is significantly slower. This slower convergence is due to the fully centralized approach, which has a considerably large action space. Generally, a larger action space necessitates more interactions with the environment to discover the optimal policy. For the fully decentralized design, the final reward does not reach the levels of AAP (CCDA) because AAP (FD) overlooks the coordination between traffic phases.

Table~\ref{tab_results_efficiency} details the performance comparisons of traffic efficiency between different methods in both synthetic and real scenarios. RL-based methods all outperform rule-based methods, namely, FT-30, FT-40, and Webster. Similar to the results in Fig.~\ref{fig_reward_curves}, when there is no interaction restriction, $\pi$-Light achieves the best performance due to its flexible actions. However, the benefits of our method become more apparent as the intervention frequency decreases. For instance, in a real-world scenario with $\Delta t = 300$, our approach surpasses 3DQN-TSCC by $58.1\%$ and $43.2\%$ at 3-way and 4-way intersections, respectively. This improvement is attributed to the inability of 3DQN-TSCC to adapt quickly to traffic changes, as it can only modify one phase at a time, leading to suboptimal performance when the intervention frequency is low. Additionally, we observe that the performance of AAP (CCDA) improves as $\Delta t$ increases in some traffic scenarios. This trend suggests that the control system only needs to keep up with the changes in traffic flow. More frequent interventions can destabilize the control system and thus degrade performance \cite{goodwin2001control}.

For the method proposed in this paper, although the AAP design allows the agent to adjust all phases at once, this approach significantly increases the action space. Table~\ref{tab_action_size} shows the action spaces for 3DQN-TSCC and different variations of the AAP method across experiment environments, highlighting that the action space for AAP (FC) is much larger than that for other methods. Even though the fully centralized AAP design (AAP (FC)) can achieve performance comparable to AAP (CCDA) in environments INT-1 and INT-3, it fails to converge in environments like INT-2, where AAP (FC) has an action space $500$ times larger than that of AAP (CCDA). On the other hand, a fully decentralized AAP design (AAP (FD)) reduces the action space but lacks effective coordination among traffic phases. The AAP (CCDA) method, introduced in this paper, not only minimizes the action space but also enhances phase coordination through its centralized critic mechanism. As shown in Table~\ref{tab_results_efficiency}, AAP (CCDA) consistently outperforms AAP (FD) across all tested environments. For instance, with $\Delta t = 120$, AAP (CCDA) reduces the average queue length by $26.9\%$ compared to AAP (FD), and by $29.9\%$ when $\Delta t = 300$.

\begin{table*}[!ht]
    \centering
    \caption{Steadiness under varying intervention frequencies with different methods on synthetic and real-world data. A lower value indicates better performance.}
    \label{tab_delta_steadiness}
    \resizebox{0.99\linewidth}{!}{
    \begin{tabular}{lccccccccc}
    \hline
     & \multicolumn{6}{c}{\textbf{Synthetic Data}} & \textbf{} & \multicolumn{2}{c}{\textbf{Real-world Data}} \\ \cline{2-7} \cline{9-10} 
     & \multicolumn{2}{c}{\textit{\textbf{INT-1}}} & \multicolumn{2}{c}{\textit{\textbf{INT-2}}} & \multicolumn{2}{c}{\textit{\textbf{INT-3}}} & \textit{\textbf{}} & \multicolumn{2}{c}{\textit{\textbf{ChenTa Road}}} \\ \cline{2-7} \cline{9-10} 
    Method & \textit{Route1-1} & \textit{Route2-1} & \textit{Route1-2} & \textit{Route2-2} & \textit{Route1-3} & \textit{Route2-3} & \textit{} & \textit{3-Way} & \textit{4-Way} \\ \hline
     & \multicolumn{9}{c}{$\Delta t = 0$} \\
    Webster & ${0.212}_{\pm0.11}$ & ${0.383}_{\pm0.2}$ & ${0.417}_{\pm0.22}$ & ${0.487}_{\pm0.27}$ & ${0.123}_{\pm0.07}$ & ${0.152}_{\pm0.1}$ &  & ${0.238}_{\pm0.16}$ & ${0.538}_{\pm0.35}$ \\
    UniTSA & ${0.335}_{\pm0.2}$ & ${0.598}_{\pm0.3}$ & ${0.691}_{\pm0.36}$ & ${0.764}_{\pm0.53}$ & ${0.202}_{\pm0.1}$ & ${0.232}_{\pm0.16}$ &  & ${0.359}_{\pm0.21}$ & ${0.763}_{\pm0.52}$ \\
    A-CATs & ${0.286}_{\pm0.16}$ & ${0.549}_{\pm0.28}$ & ${0.651}_{\pm0.36}$ & ${0.594}_{\pm0.34}$ & ${0.204}_{\pm0.12}$ & ${0.225}_{\pm0.15}$ &  & ${0.310}_{\pm0.17}$ & ${0.660}_{\pm0.42}$ \\
    3DQN-TSCC & ${0.052}_{\pm0.03}$ & ${0.103}_{\pm0.07}$ & ${0.113}_{\pm0.06}$ & ${0.181}_{\pm0.11}$ & ${0.039}_{\pm0.02}$ & ${0.064}_{\pm0.04}$ &  & ${0.079}_{\pm0.06}$ & ${0.173}_{\pm0.10}$ \\
    AAP (FC) & ${0.150}_{\pm0.08}$ & ${0.320}_{\pm0.22}$ & ${0.333}_{\pm0.17}$ & ${0.497}_{\pm0.29}$ & ${0.131}_{\pm0.07}$ & ${0.141}_{\pm0.07}$ &  & ${0.158}_{\pm0.1}$ & ${0.513}_{\pm0.31}$ \\
    AAP (FD) & ${0.166}_{\pm0.09}$ & ${0.372}_{\pm0.19}$ & ${0.374}_{\pm0.25}$ & ${0.630}_{\pm0.43}$ & ${0.139}_{\pm0.07}$ & ${0.179}_{\pm0.1}$ &  & ${0.194}_{\pm0.13}$ & ${0.546}_{\pm0.29}$ \\
    AAP (CCDA) & ${0.177}_{\pm0.12}$ & ${0.370}_{\pm0.25}$ & ${0.356}_{\pm0.18}$ & ${0.654}_{\pm0.35}$ & ${0.140}_{\pm0.09}$ & ${0.175}_{\pm0.11}$ &  & ${0.202}_{\pm0.14}$ & ${0.575}_{\pm0.3}$ \\ \hline
     & \multicolumn{9}{c}{$\Delta t = 60$} \\
    3DQN-TSCC & ${0.033}_{\pm0.02}$ & ${0.065}_{\pm0.04}$ & ${0.071}_{\pm0.05}$ & ${0.114}_{\pm0.06}$ & ${0.023}_{\pm0.01}$ & ${0.044}_{\pm0.03}$ &  & ${0.068}_{\pm0.04}$ & ${0.099}_{\pm0.06}$ \\
    AAP (FC) & ${0.155}_{\pm0.08}$ & ${0.316}_{\pm0.2}$ & ${0.323}_{\pm0.21}$ & ${0.480}_{\pm0.29}$ & ${0.104}_{\pm0.06}$ & ${0.179}_{\pm0.09}$ &  & ${0.136}_{\pm0.07}$ & ${0.499}_{\pm0.3}$ \\
    AAP (FD) & ${0.139}_{\pm0.09}$ & ${0.262}_{\pm0.15}$ & ${0.317}_{\pm0.17}$ & ${0.458}_{\pm0.31}$ & ${0.085}_{\pm0.06}$ & ${0.178}_{\pm0.11}$ &  & ${0.163}_{\pm0.10}$ & ${0.401}_{\pm0.21}$ \\
    AAP (CCDA) & ${0.162}_{\pm0.11}$ & ${0.318}_{\pm0.19}$ & ${0.325}_{\pm0.2}$ & ${0.578}_{\pm0.39}$ & ${0.100}_{\pm0.06}$ & ${0.164}_{\pm0.11}$ &  & ${0.201}_{\pm0.13}$ & ${0.500}_{\pm0.37}$ \\ \hline
     & \multicolumn{9}{c}{$\Delta t = 120$} \\
    3DQN-TSCC & ${0.017}_{\pm0.01}$ & ${0.037}_{\pm0.02}$ & ${0.037}_{\pm0.02}$ & ${0.062}_{\pm0.03}$ & ${0.014}_{\pm0.01}$ & ${0.023}_{\pm0.01}$ &  & ${0.039}_{\pm0.03}$ & ${0.051}_{\pm0.03}$ \\
    AAP (FC) & ${0.055}_{\pm0.04}$ & ${0.127}_{\pm0.07}$ & ${0.120}_{\pm0.08}$ & ${0.210}_{\pm0.13}$ & ${0.049}_{\pm0.03}$ & ${0.078}_{\pm0.05}$ &  & ${0.130}_{\pm0.07}$ & ${0.171}_{\pm0.11}$ \\
    AAP (FD) & ${0.054}_{\pm0.03}$ & ${0.131}_{\pm0.09}$ & ${0.131}_{\pm0.08}$ & ${0.221}_{\pm0.11}$ & ${0.049}_{\pm0.03}$ & ${0.079}_{\pm0.05}$ &  & ${0.133}_{\pm0.07}$ & ${0.171}_{\pm0.09}$ \\
    AAP (CCDA) & ${0.052}_{\pm0.03}$ & ${0.132}_{\pm0.07}$ & ${0.141}_{\pm0.08}$ & ${0.235}_{\pm0.12}$ & ${0.050}_{\pm0.03}$ & ${0.085}_{\pm0.05}$ &  & ${0.125}_{\pm0.09}$ & ${0.164}_{\pm0.11}$ \\ \hline
     & \multicolumn{9}{c}{$\Delta t = 300$} \\
    3DQN-TSCC & ${0.007}_{\pm0.0}$ & ${0.016}_{\pm0.01}$ & ${0.014}_{\pm0.01}$ & ${0.028}_{\pm0.02}$ & ${0.005}_{\pm0.0}$ & ${0.010}_{\pm0.01}$ &  & ${0.015}_{\pm0.01}$ & ${0.023}_{\pm0.02}$ \\
    AAP (FC) & ${0.021}_{\pm0.01}$ & ${0.045}_{\pm0.02}$ & ${0.038}_{\pm0.02}$ & ${0.078}_{\pm0.04}$ & ${0.017}_{\pm0.01}$ & ${0.027}_{\pm0.01}$ &  & ${0.026}_{\pm0.02}$ & ${0.039}_{\pm0.04}$ \\
    AAP (FD) & ${0.017}_{\pm0.01}$ & ${0.041}_{\pm0.03}$ & ${0.035}_{\pm0.02}$ & ${0.071}_{\pm0.04}$ & ${0.014}_{\pm0.01}$ & ${0.025}_{\pm0.01}$ &  & ${0.034}_{\pm0.03}$ & ${0.054}_{\pm0.04}$ \\
    AAP (CCDA) & ${0.015}_{\pm0.01}$ & ${0.036}_{\pm0.02}$ & ${0.031}_{\pm0.02}$ & ${0.057}_{\pm0.04}$ & ${0.012}_{\pm0.01}$ & ${0.020}_{\pm0.01}$ &  & ${0.028}_{\pm0.03}$ & ${0.030}_{\pm0.04}$ \\ \hline
    \end{tabular}}
\end{table*}

\begin{figure*}[!ht]
    \centering
    \subfloat[]{\includegraphics[width=0.48\textwidth]{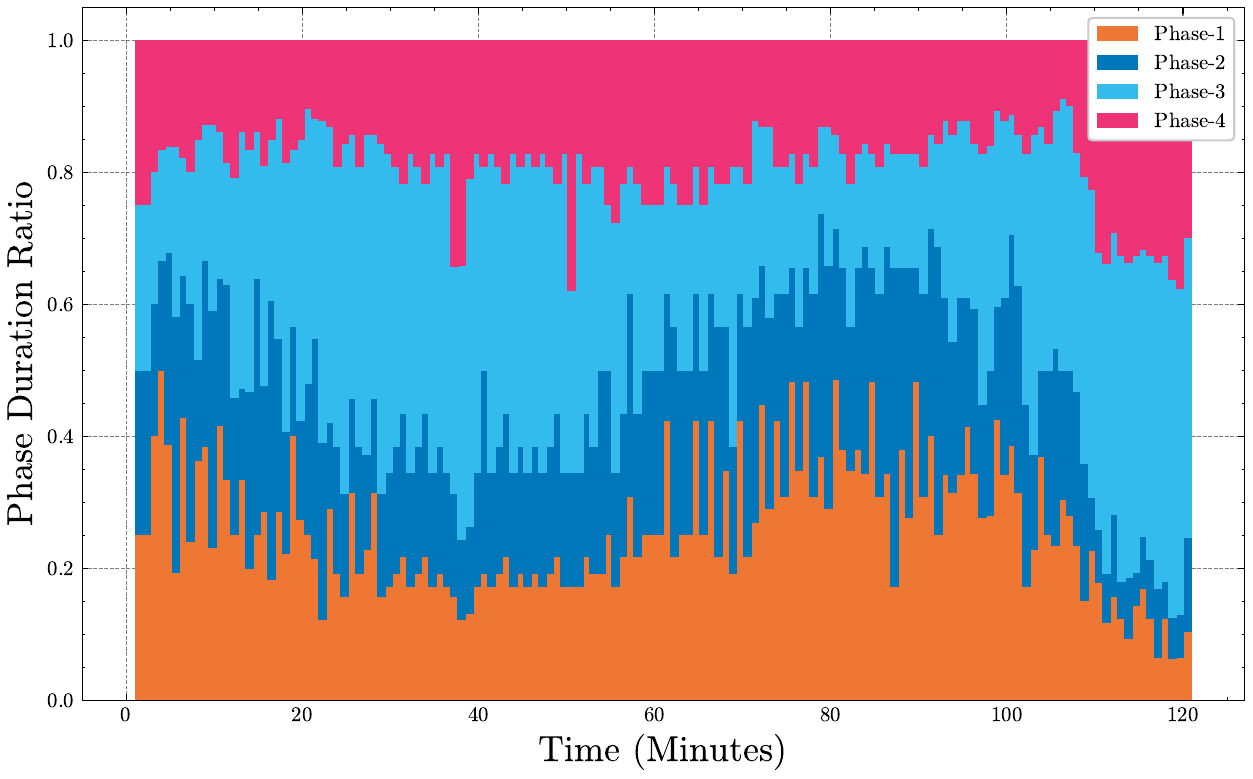}
    \label{fig_steadiness_None}}
    \subfloat[]{\includegraphics[width=0.48\textwidth]{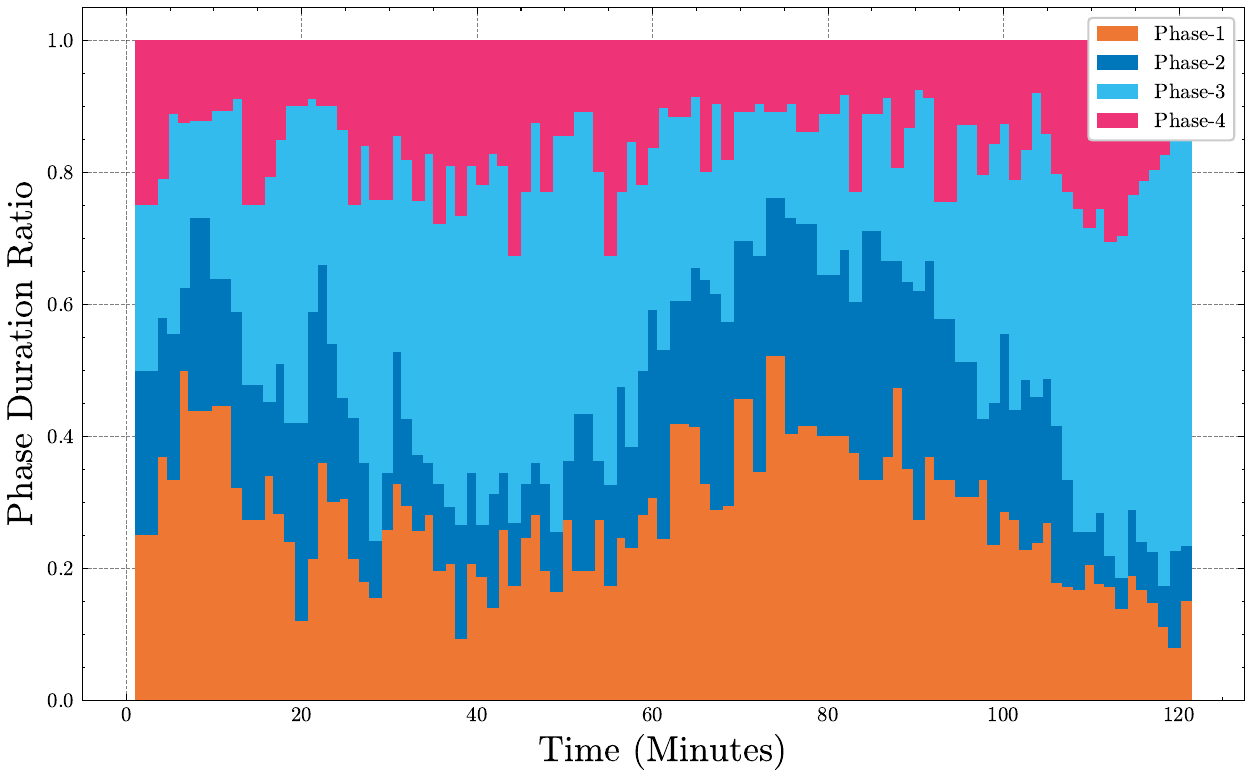}
    \label{fig_steadiness_60}}
    
    \hfill
    
    \subfloat[]{\includegraphics[width=0.48\textwidth]{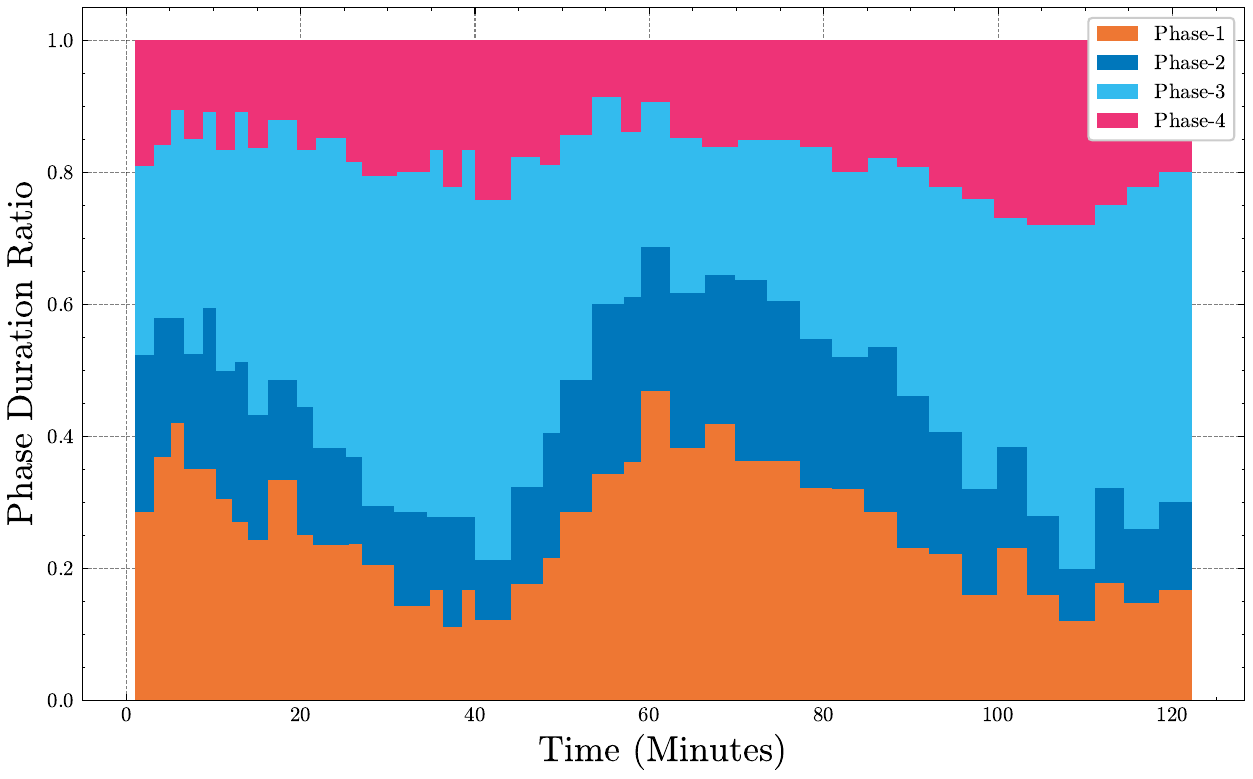}
    \label{fig_steadiness_120}}
    \subfloat[]{\includegraphics[width=0.48\textwidth]{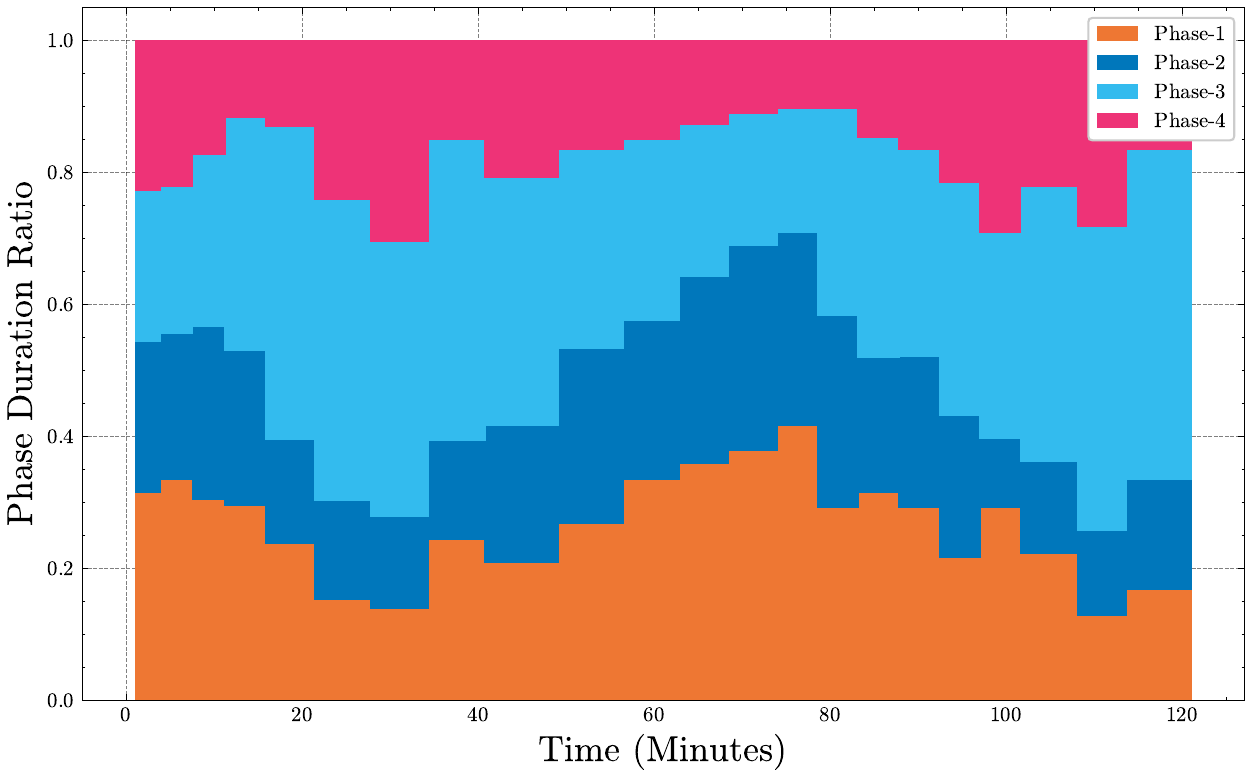}
    \label{fig_steadiness_300}}
    
    \caption{Variation in Phase Durations of AAP (CCDA) at Different Control Frequencies in the INT-1, Route2-1 Environment. (a) $\Delta t=0$. (b) $\Delta t=60$. (c) $\Delta t=120$. (d) $\Delta t=300$.}
    \label{fig_steadiness_example}
\end{figure*}

\begin{figure*}[!ht]
    \centering
    \subfloat[]{\includegraphics[width=0.32\textwidth]{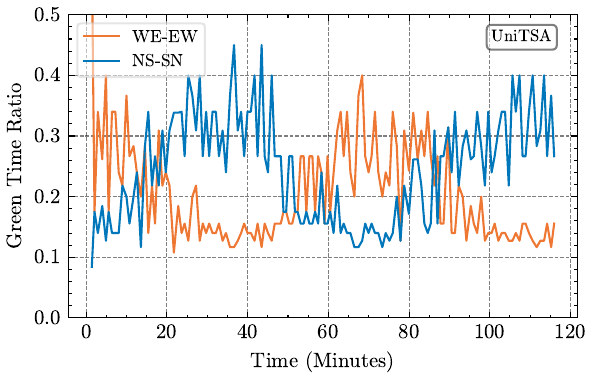}
    \label{fig:syn_green_a}}
    % \hfil
    \subfloat[]{\includegraphics[width=0.32\textwidth]{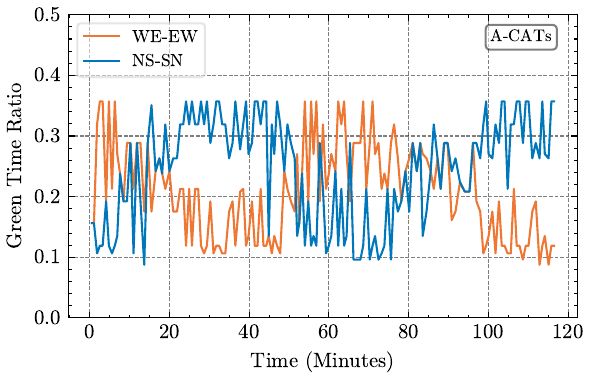}
    \label{fig:syn_green_b}}
    % \hfil
    \subfloat[]{\includegraphics[width=0.32\textwidth]{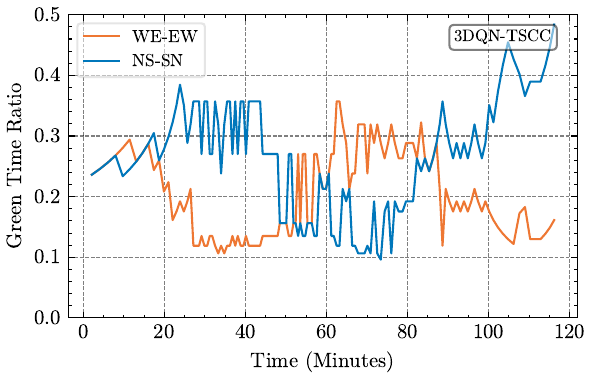}
    \label{fig:syn_green_c}}
    
    \hfill
    
    \subfloat[]{\includegraphics[width=0.32\textwidth]{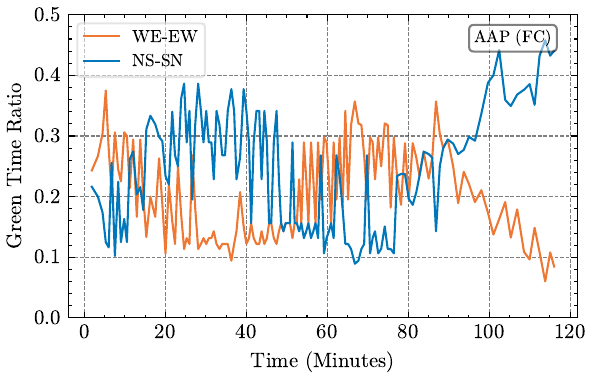}
    \label{fig:syn_green_d}}
    \subfloat[]{\includegraphics[width=0.32\textwidth]{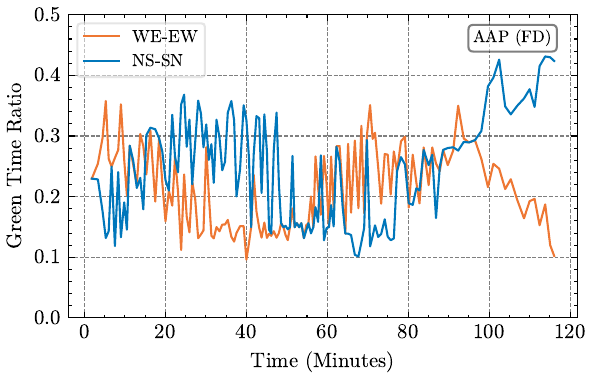}
    \label{fig:syn_green_e}}
    \subfloat[]{\includegraphics[width=0.32\textwidth]{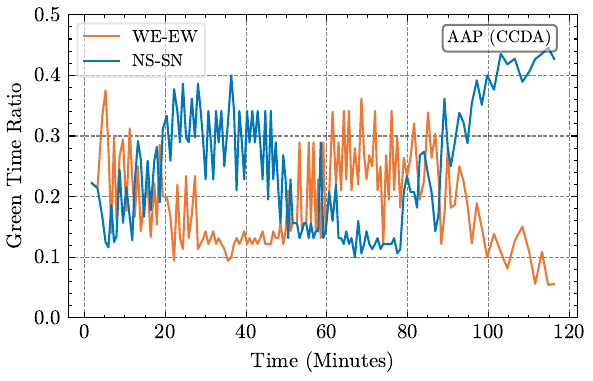}
    \label{fig:syn_green_f}}
    \caption{Green time ratio comparison under INT-1, Route2-1. (a) UniTSA. (b) A-CATs. (c) 3DQN-TSCC. (d) AAP (FC). (e) AAP (FD). (f) AAP (CCDA).}
    \label{fig:syn_green}
\end{figure*}

\begin{figure*}[!ht]
    \centering
    \subfloat[]{\includegraphics[width=0.32\textwidth]{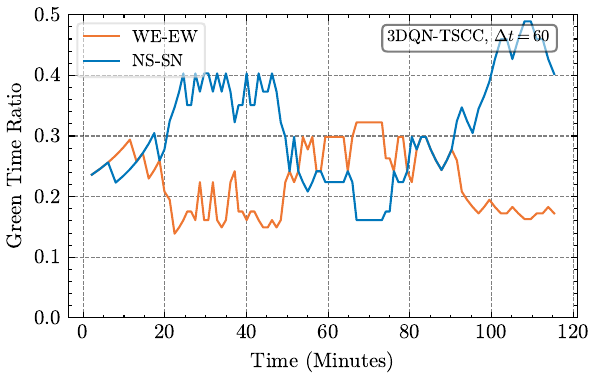}
    \label{fig_asp_60}}
    \subfloat[]{\includegraphics[width=0.32\textwidth]{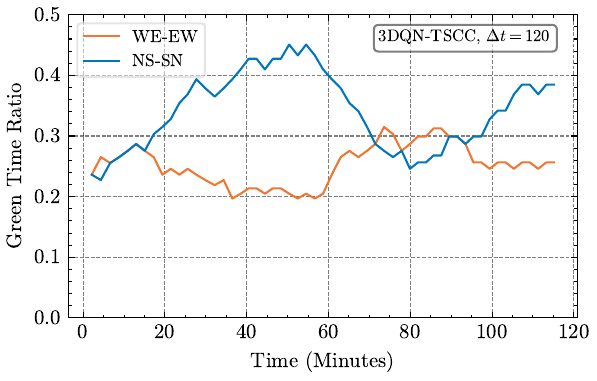}
    \label{fig_asp_120}}
    \subfloat[]{\includegraphics[width=0.32\textwidth]{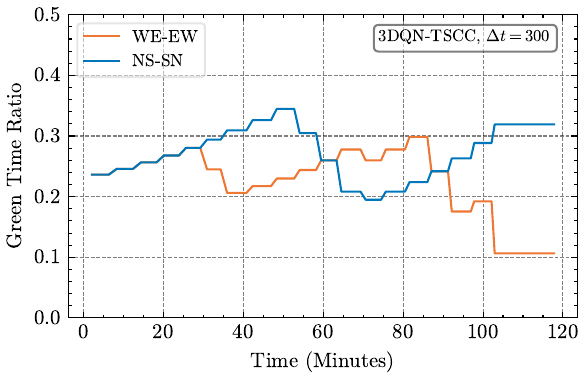}
    \label{fig_asp_300}} 
    
    \hfill
    
    \subfloat[]{\includegraphics[width=0.32\textwidth]{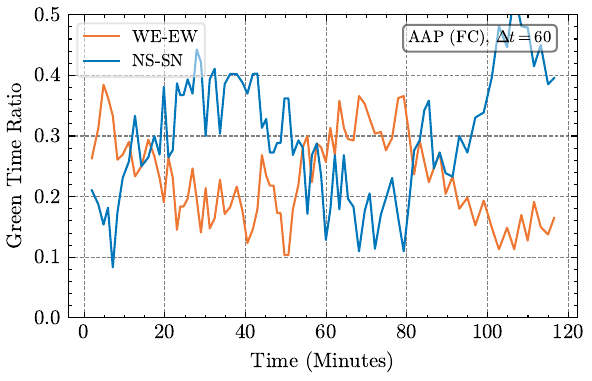}
    \label{fig_aap_fc_60}}
    \subfloat[]{\includegraphics[width=0.32\textwidth]{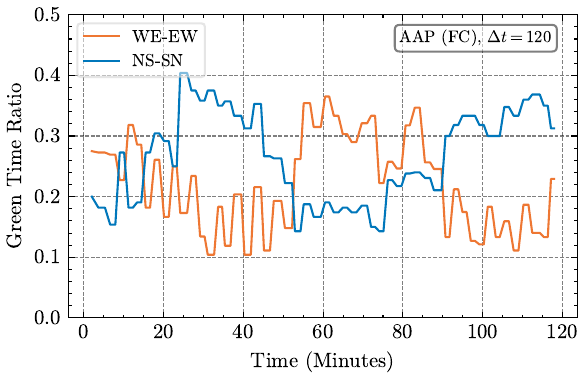}
    \label{fig_aap_fc_120}}
    \subfloat[]{\includegraphics[width=0.32\textwidth]{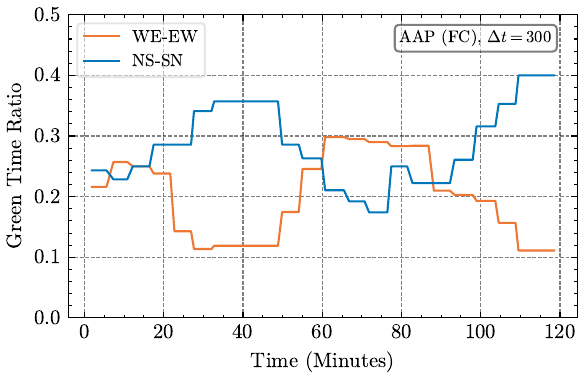}
    \label{fig_aap_fc_300}}

    \hfill

    \subfloat[]{\includegraphics[width=0.32\textwidth]{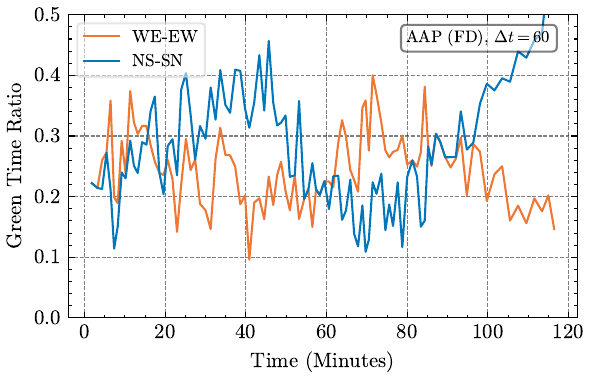}
    \label{fig_aap_fd_60}}
    \subfloat[]{\includegraphics[width=0.32\textwidth]{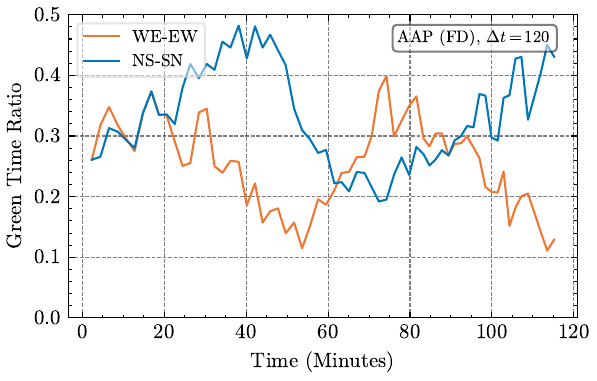}
    \label{fig_aap_fd_120}}
    \subfloat[]{\includegraphics[width=0.32\textwidth]{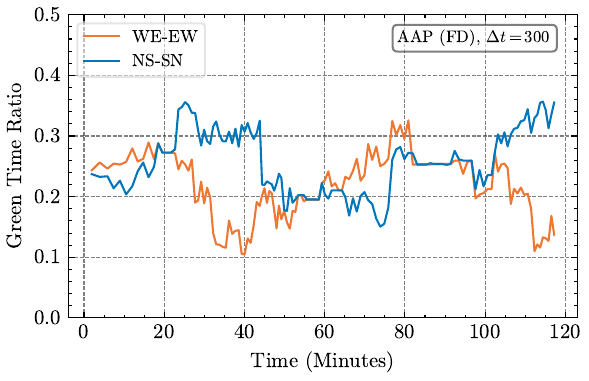}
    \label{fig_aap_fd_300}}

    \hfill

    \subfloat[]{\includegraphics[width=0.32\textwidth]{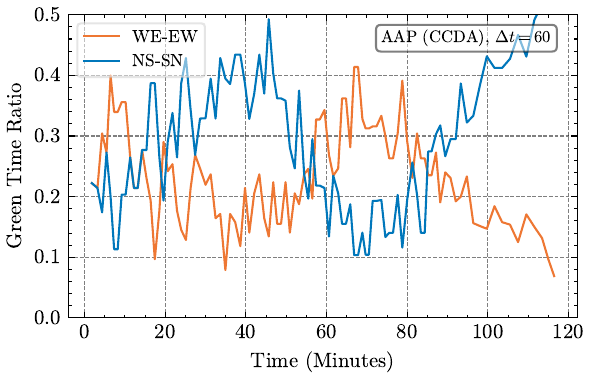}
    \label{fig_aap_ccda_60}}
    \subfloat[]{\includegraphics[width=0.32\textwidth]{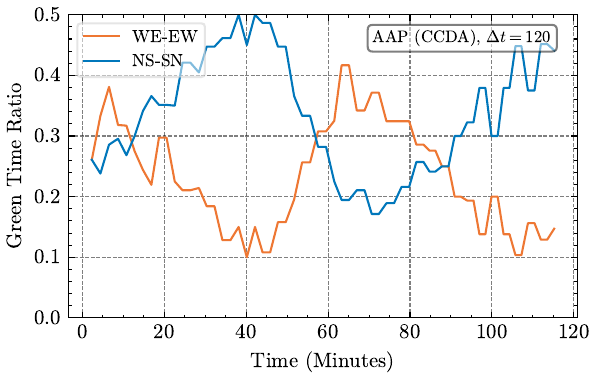}
    \label{fig_aap_ccda_120}}
    \subfloat[]{\includegraphics[width=0.32\textwidth]{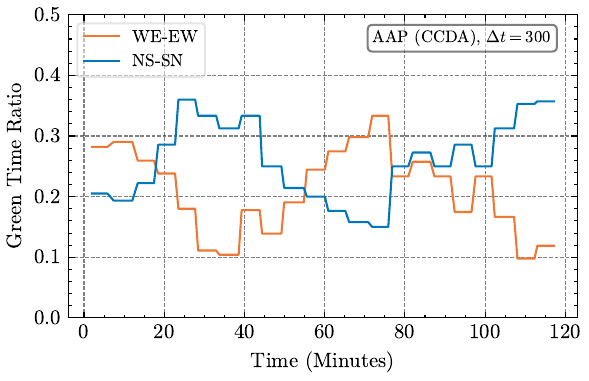}
    \label{fig_aap_ccda_300}}
    
    \caption{Green time ratios for various traffic control algorithms at different intervention frequencies. (a) to (c) illustrate 3DQN-TSCC, (d) to (f) AAP (FC), (g) to (i) AAP (FD), and (j) to (l) AAP (CCDA), each set with $\Delta t$ of 60, 120, and 300, respectively.}
    \label{fig_syn_delta_green_ratio}
\end{figure*}

\begin{figure*}[!ht]
  \centering
  \subfloat[]{\includegraphics[width=0.49\textwidth]{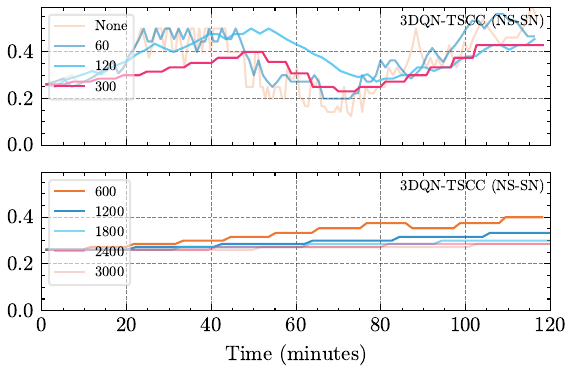}
  \label{fig_3dqntscc_NS-SN}}
  \subfloat[]{\includegraphics[width=0.49\textwidth]{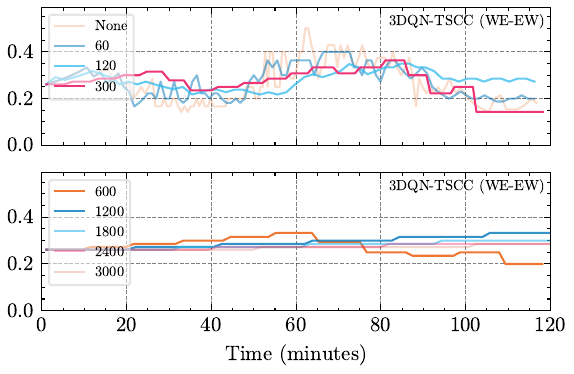}
  \label{fig_3dqntscc_WE-EW}}

  \subfloat[]{\includegraphics[width=0.49\textwidth]{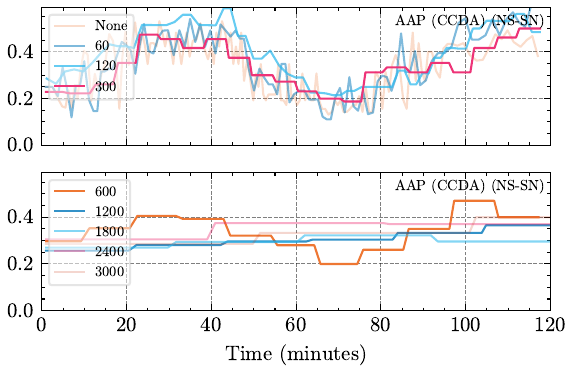}
  \label{fig_aap_ccda_NS-SN}}
  \subfloat[]{\includegraphics[width=0.49\textwidth]{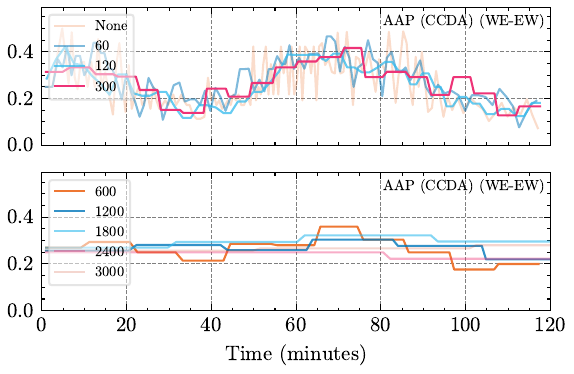}
  \label{fig_aap_ccda_WE-EW}}  
  \caption{Variation of green time ratios with different $\Delta t$ values for 3DQN-TSCC and AAP (CCDA) in NS-SN and WE-EW Directions. (a) 3DQN-TSCC, NS-SN. (b) 3DQN-TSCC, WE-EW. (c) APP (CCDA), NS-SN. (d) APP (CCDA), WE-EW.}
  \label{fig_syn_green_ratio_delta}
\end{figure*}

\begin{figure}[!ht]
  \centering
  \includegraphics[width=0.6\linewidth]{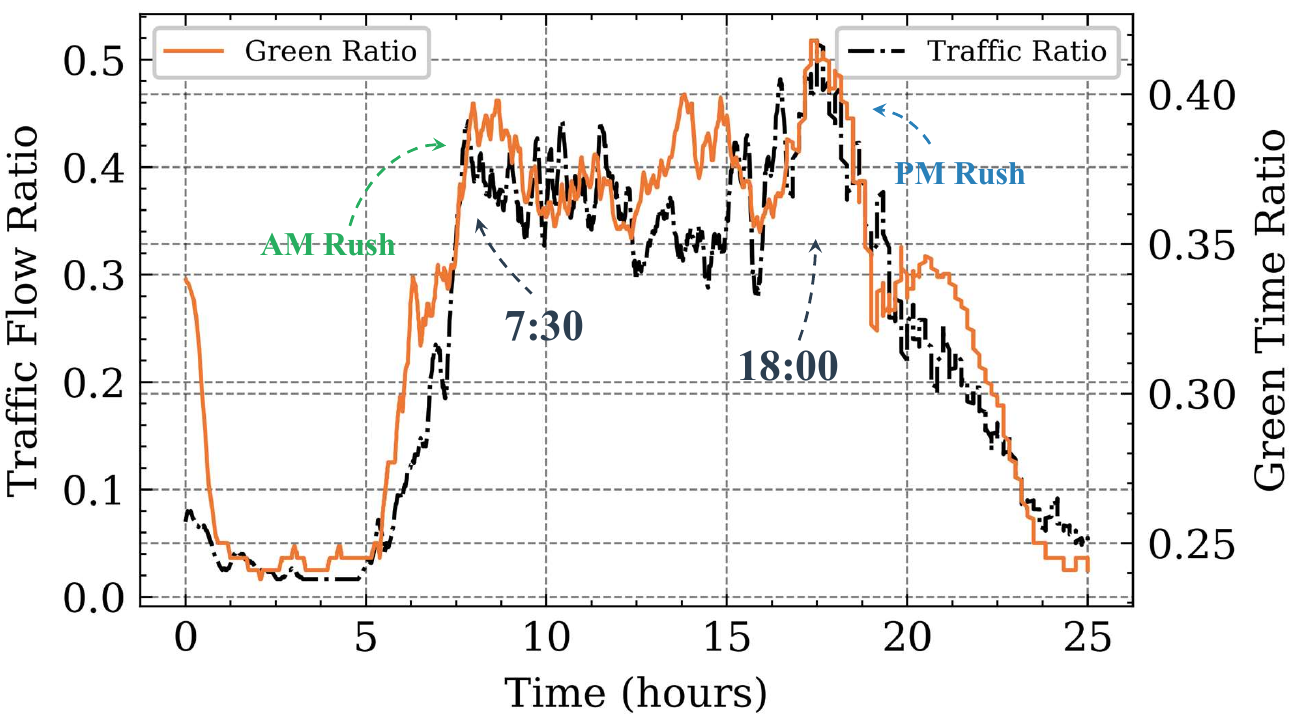}
  \caption{Average arrival rate and green time ratio from AAP (CCDA) in real-world data under $\Delta t = 300$.} 
  \label{fig_real_data_ratio_arrival_rate}
\end{figure}

\subsection{Comparison of Steadiness}

In addition to traffic efficiency, the steadiness of a traffic control policy is a critical consideration in real-world applications. Table~\ref{tab_delta_steadiness} presents the average steadiness scores, where lower values indicate smaller changes and thus greater stability. These results cover both synthetic and real-world scenarios under various intervention frequencies. When the action frequency is unrestricted $\Delta t = 0$, both UniTSA and A-CATs exhibit low steadiness scores. This outcome occurs because these methods adjust the traffic phase directly without limiting the magnitude of the change in phase timing. Thus, the duration of consecutive green light time can be very different. In contrast, the 3DQN-TSCC method demonstrates superior stability across all methods by modifying each phase only once per cycle. The AAP-based methods introduced in this paper, while being slightly less stable than 3DQN-TSCC, outperform other benchmark methods. This difference stems from AAP's ability to adjust all traffic phases within one cycle to accommodate complex and dynamic traffic conditions, striking a balance between efficiency and steadiness.

As the intervention frequency decreases, the stability scores for all methods improve, and the performance gap between our AAP (CCDA) method and 3DQN-TSCC narrows. Specifically, with no intervention restrictions, the stability score of AAP (CCDA) is 0.2 higher than that of 3DQN-TSCC in the ChenTa Road scenario, suggesting greater instability. However, when interventions occur every 300 seconds, the stability score of AAP (CCDA) at ChenTa intersections is only 0.01 higher than that of 3DQN-TSCC, marking a tenfold reduction in the difference. Despite this, as indicated in Table~\ref{tab_results_efficiency}, AAP (CCDA) achieved a $51\%$ higher traffic efficiency compared to 3DQN-TSCC at the same intervention frequency. This indicates that AAP (CCDA) can enhance stability while maintaining high traffic efficiency, even at lower intervention frequencies. Additionally, when $\Delta t = 300$, all stability values are below $0.05$, indicating generally stable traffic signal timing with relatively minor adjustments in the acceleration or deceleration of phase duration changes. It can also be observed that reducing the frequency of interventions leads to fewer changes over short periods, contributing to a more stable and predictable traffic flow.

Fig.~\ref{fig_steadiness_example} illustrates the variation in the duration of the four phases of AAP (CCDA) at different action intervals in the INT-1, Route2-1 environment. The x-axis denotes time, while the y-axis represents the duration ratio of each phase. The figure shows frequent changes in phase duration at $\Delta t = 0$, evident from numerous spikes in Fig.~\ref{fig_steadiness_None}. However, as the control frequency decreases, the change in phase duration becomes smoother, and the number of spikes reduces. Notably, at a control frequency of $300$~s, as depicted in Fig.~\ref{fig_steadiness_300}, each phase lasts for five minutes, with transitions occurring less abruptly, resulting in smoother transitions.

\subsection{Interpretation of Learned Policy} \label{sec_interpretation_learned_policy}

In this section, we compare signal policies learned by benchmarking methods and the proposed AAP (CCDA) by analyzing traffic flow and green time ratio under INT-1 and Route2-1. Analyzing these strategies is crucial as it helps understand why AAP (CCDA) performs well even at low intervention frequencies. Traffic flow in INT-1 and Route2-1 is illustrated in Fig.~\ref{fig_traffic_flow}, which shows fluctuations throughout the entire experiment. Fig.~\ref{fig:syn_green} displays the policies learned by different RL-based TSC methods when $\Delta t = 0$. The trends in the green time ratio generally correlate with the fluctuations in traffic flow. This indicates that under high-frequency interaction environments, all methods can learn effective strategies, thus ensuring good traffic efficiency. Among these methods, 3DQN-TSCC was notably more steady than the others, as shown in Fig.~\ref{fig:syn_green_c}, consistent with its favorable steadiness score in Table~\ref{tab_delta_steadiness}. 

To further understand the impact of different intervention frequencies on the traffic signal control strategies, we analyze the results of 3DQN-TSCC and the methods proposed in this paper at intervention frequencies of $\Delta t = 60, 120, 300$. Fig.~\ref{fig_syn_delta_green_ratio} shows the change in the green time ratio in INT-1 and Route2-1 under these intervention frequencies. It is evident that the green time ratios for these methods become smoother as the intervention frequency decreases. With 3DQN-TSCC, limited by intervention frequency, the strategies cannot keep up with changes in traffic flow. For example, when $\Delta t = 300$, traffic in the NS-SN direction begins to decrease at 40 minutes, but the policy adjustment by 3DQN-TSCC is delayed until 50 minutes because it can only modify one traffic phase at a time. In contrast, the AAP-based methods introduced in this study can adjust all traffic phases simultaneously. Both AAP (FC) and AAP (CCDA) incorporate cooperation among traffic phases, allowing their strategies to better align with traffic flow dynamics. Although AAP (FC) converges slowly, the results after convergence also align well with the trend of traffic flow. For AAP (FD), since each traffic phase is adjusted independently, it is apparent that the green ratios of NS-SN and WE-EW cannot be well coordinated, further supporting the performance mentioned in Table~\ref{tab_results_efficiency}.

As illustrated in Fig.~\ref{fig_aap_ccda_300}, the strategy learned by AAP (CCDA) remains effective even with a control frequency as low as $5$ minutes. To determine the minimum effective frequency for AAP (CCDA), experiments were also conducted on INT-1 and Route2-1 with varying $\Delta t$ from $0$ to $2000$ seconds. Fig.~\ref{fig_syn_green_ratio_delta} displays the changes in the green time ratio for the NS-SN and WE-EW directions as $\Delta t$ varies, comparing 3DQN-TSCC and AAP (CCDA). Observations from Fig.~\ref{fig_3dqntscc_NS-SN} and Fig.~\ref{fig_3dqntscc_WE-EW} indicate that for $\Delta t < 300$, the green time ratio under 3DQN-TSCC closely follows the pattern observed without any delay. However, beyond this threshold, 3DQN-TSCC struggles to adapt to fluctuations in traffic flow. In contrast, as depicted in Fig.~\ref{fig_aap_ccda_NS-SN} and Fig.~\ref{fig_aap_ccda_WE-EW}, AAP (CCDA) maintains robust performance even at larger $\Delta t$ values. Notably, with $\Delta t$ set to $600$ (approximately 10 minutes), the green time ratio effectively aligns with the traffic flow variations, underscoring the advantage of AAP (CCDA) in scenarios requiring low-frequency interventions.

We further analyze the policy learned from the AAP (CCDA) using real-world data at low intervention frequencies. Specifically, we focus on the traffic signal policy at the intersection of Chenta Road in the SN-NS directions, comparing peak and non-peak hours. Fig.~\ref{fig_real_data_ratio_arrival_rate} presents the average arrival rate and the green time ratio for the SN-NS phase at this 4-way intersection with $\Delta t = 300$. The data illustrates that the green time ratio adapts to the fluctuations in traffic flow. Notably, during peak hours, around 7:30 AM and 6:00 PM, the policy extends the green light duration for the NS-SN direction more than during non-peak hours. This adaptation indicates that the proposed AAP (CCDA) method effectively adjusts to complex traffic patterns even with infrequent interventions.

\begin{table}[!ht]
    \centering
    \caption{Efficiency of Different Magnitudes of Change in Traffic Phase Duration.}
    \label{tab_sensitivity_analysis_efficiency}
    \begin{tabular}{lccccc}
    \hline
    \multicolumn{1}{c}{} & \multicolumn{5}{c}{\textbf{ChenTa Road}} \\ \cline{2-6} 
    \multirow{2}{*}{Magnitude of Change} & \multicolumn{2}{c}{$\Delta t = 0$} &  & \multicolumn{2}{c}{$\Delta t = 300$} \\ \cline{2-3} \cline{5-6} 
    \multicolumn{1}{c}{} & 3-Way & 4-Way &  & 3-Way & 4-Way \\ \cline{1-3} \cline{5-6} 
    \{-5,0,5\} & $0.812$ & $5.645$ &  & $1.811$ & $8.207$ \\
    \{-6,-3,0,3,6\} & $0.678$ & $5.393$ &  & $1.813$ & $7.679$ \\
    \{-10,-5,0,5,10\} & $0.795$ & $5.809$ &  & $1.828$ & $8.435$ \\ \hline
    \end{tabular}
\end{table}

\begin{table}[!ht]
    \centering
    \caption{Steadiness of Different Magnitudes of Change in Traffic Phase Duration.}
    \label{tab_sensitivity_analysis_steadiness}
    \begin{tabular}{lccccc}
    \hline
    \multicolumn{1}{c}{} & \multicolumn{5}{c}{\textbf{ChenTa Road}} \\ \cline{2-6} 
    \multirow{2}{*}{Magnitude of Change} & \multicolumn{2}{c}{$\Delta t = 0$} &  & \multicolumn{2}{c}{$\Delta t = 300$} \\ \cline{2-3} \cline{5-6} 
    \multicolumn{1}{c}{} & 3-Way & 4-Way &  & 3-Way & 4-Way \\ \cline{1-3} \cline{5-6} 
    \{-5,0,5\} & $0.170$ & $0.437$ &  & $0.024$ & $0.029$ \\
    \{-6,-3,0,3,6\} & $0.202$ & $0.575$ &  & $0.028$ & $0.030$ \\
    \{-10,-5,0,5,10\} & $0.217$ & $0.545$ &  & $0.025$ & $0.031$ \\ \hline
    \end{tabular}
\end{table}

\subsection{Sensitivity Analysis under Different Magnitudes of Change} \label{sec_sensitivity_analysis}

In this section, we conduct a sensitivity analysis to evaluate the impact of varying magnitudes of change on the efficiency and steadiness of our proposed AAP (CCDA) approach at ChenTa Road. We consider three sets of magnitude changes: $\{-5, 0, 5\}$, $\{-6, -3, 0, 3, 6\}$, and $\{-10, -5, 0, 5, 10\}$, which correspond to small, moderate, and large variations, respectively. The efficiency results are shown in Table~\ref{tab_sensitivity_analysis_efficiency} and the steadiness results in Table~\ref{tab_sensitivity_analysis_steadiness}. Measurements were taken under two conditions: without any intervention ($\Delta t = 0$) and with interventions occurring every 300 seconds ($\Delta t = 300$).

Table~\ref{tab_sensitivity_analysis_efficiency} reveals that the set $\{-6, -3, 0, 3, 6\}$ generally achieves the shortest queue lengths. This effectiveness can be attributed to the inclusion of both substantial adjustments $-6$ and $6$ and smaller changes $-3$ and $3$. It is important to note that the impact of the different magnitudes of change on performance does not exceed a variation of $10\%$, and all variations outperform the 3DQN-TSCC method.

From Table~\ref{tab_sensitivity_analysis_steadiness}, it is observed that the set $\{-5, 0, 5\}$ exhibits the best stability, attributed to its narrower range of adjustments. Additionally, as the intervention interval increases, the methods with various magnitudes of change show enhanced stability while maintaining efficiency. These results demonstrate that the AAP (CCDA) approach can adapt effectively to different magnitudes of change, offering a versatile solution for traffic signal control that can be tailored to specific traffic conditions.

% %%%%%%%%%%
% Conclusion
% %%%%%%%%%%
\section{Conclusion} \label{sec_conclusion}

In this study, we developed a traffic signal control system designed to operate efficiently across different intervention frequencies. Our proposed method, named AAP (CCDA), introduces a novel approach that allows for simultaneous adjustments of all traffic signal phases within a single cycle. This unique action design ensures short-term stability and maintains high performance even at reduced frequencies of intervention. The AAP (CCDA) method addresses the challenges of an expanding action space due to this novel action design by employing a RL framework with decentralized actors and a centralized critic. The decentralized actors reduce complexity by enabling the RL agent to output adjustments for each individual traffic phase, rather than managing combinations of all phases, while the centralized critic provides comprehensive evaluations to enhance cooperation among traffic phases. Extensive testing on both simulated and real-world traffic scenarios has shown that the AAP (CCDA) method can decrease the average queue length by up to $58.1\%$ compared to traditional traffic signal control methods, particularly under low intervention frequencies. Both quantitative and qualitative evaluations confirm the robustness and efficiency of the AAP (CCDA) system under varying control frequencies.

Looking ahead, several promising directions could further enhance the CCDA framework. A method needs to be designed to determine the optimal frequency selection in TSC systems and to quantitatively measure the impacts of different intervention frequencies on resource utilization, safety, and traffic flow disruptions. Additionally, it is necessary to design the TSC system to adapt to dynamic intervention frequencies instead of a fixed control interval, as the intervention frequency may vary according to the actual traffic flow. Extending this framework to manage thousands of urban intersections simultaneously presents a significant opportunity to improve city-wide traffic flow and effectively reduce congestion.

% %%%%%%%%%%%%%%%%
% Acknowledgements
% %%%%%%%%%%%%%%%%
\section{Acknowledgements}

This work was supported in part by the Shenzhen Science and Technology Innovation Committee under Grant No. JCYJ20190813170803617 and the Shanghai Pujiang Program under Grant No.21PJD092. Work at the Stanford Center at the Incheon Global Campus (SCIGC) is supported in part under the National Program to Subsidize Attracting Foreign Educational Institution and Research Institutes published by the Ministry of Trade, Industry, and Energy of the Republic of Korea and managed by the Incheon Free Economic Zone Authority.

% %%%%%%%%%
% Reference
% %%%%%%%%%
\bibliographystyle{IEEEtran}
\bibliography{ref}

\end{document}